\pdfoutput=1
\documentclass{article}

\usepackage{PRIMEarxiv}

\usepackage[utf8]{inputenc}
\usepackage[T1]{fontenc}
\usepackage{booktabs}
\usepackage{array}
\usepackage{tabularx}
\usepackage{longtable}
\usepackage{float}
\usepackage{microtype}
\usepackage{graphicx}
\usepackage{fancyhdr}
\usepackage{xspace}
\xspaceaddexceptions{\%}
\usepackage[round]{natbib}
\usepackage{url}
\usepackage{xurl}
\usepackage{hyperref}
\hypersetup{colorlinks=true, linkcolor=blue, citecolor=blue, urlcolor=blue}
\hypersetup{pdftitle={Before They Were Cool: Do Early Fans Abandon Artists Who Suddenly Become Popular? Evidence from Online Music Listening},
  pdfauthor={Zihan Chen, Lei Nico Zheng, Di Zhu},
  pdfkeywords={identity signaling, taste performance, popularity, early adopters, music listening, self-presentation}}
\graphicspath{{figures/}}

\newcommand{\AuditBatchC}{70\xspace}
\newcommand{\AuditBatchT}{14\xspace}
\newcommand{\AuditCleanNoCtrl}{4\xspace}
\newcommand{\AuditCleanT}{31\xspace}
\newcommand{\AuditEvents}{27\xspace}
\newcommand{\AuditFlagPctC}{60\xspace}
\newcommand{\AuditFlagPctT}{60\xspace}
\newcommand{\AuditFlaggedC}{213\xspace}
\newcommand{\AuditFlaggedT}{47\xspace}
\newcommand{\AuditNewC}{149\xspace}
\newcommand{\AuditNewPctC}{42\xspace}
\newcommand{\AuditNewPctT}{51\xspace}
\newcommand{\AuditNewT}{40\xspace}
\newcommand{\AuditQuit}{\ensuremath{-}3.4\xspace}
\newcommand{\AuditQuithi}{\ensuremath{-}1.2\xspace}
\newcommand{\AuditQuitlo}{\ensuremath{-}5.4\xspace}
\newcommand{\AuditUncC}{64\xspace}
\newcommand{\AuditUncT}{7\xspace}
\newcommand{\AuditUnitsC}{357\xspace}
\newcommand{\AwCeremonies}{9\xspace}
\newcommand{\AwLisGrowthC}{75\xspace}
\newcommand{\AwLisGrowthT}{81\xspace}
\newcommand{\AwLooShareMax}{0.020\xspace}
\newcommand{\AwLooShareMin}{\ensuremath{-}0.127\xspace}
\newcommand{\AwNevents}{18\xspace}
\newcommand{\AwNewRatioC}{1.0\xspace}
\newcommand{\AwNewRatioT}{1.2\xspace}
\newcommand{\AwPermSharePct}{70\xspace}
\newcommand{\AwQuitDiff}{\ensuremath{-}2.1\xspace}
\newcommand{\AwQuitDiffhi}{0.3\xspace}
\newcommand{\AwQuitDifflo}{\ensuremath{-}4.9\xspace}
\newcommand{\AwQuitPlaceboDiff}{\ensuremath{-}1.7\xspace}
\newcommand{\AwReturnDiff}{2.5\xspace}
\newcommand{\AwReturnPlaceboDiff}{4.3\xspace}
\newcommand{\AwSharePctloAbs}{23.7\xspace}
\newcommand{\AwSharePost}{\ensuremath{-}0.035\xspace}
\newcommand{\AwSharePosthi}{0.186\xspace}
\newcommand{\AwSharePostlo}{\ensuremath{-}0.269\xspace}
\newcommand{\AwShareRef}{1.14\xspace}
\newcommand{\AwWinnerArtists}{17\xspace}
\newcommand{\CatAlbumPct}{83.1\xspace}
\newcommand{\CatOtherArtistPct}{5.7\xspace}
\newcommand{\CatOwnOtherPct}{11.2\xspace}
\newcommand{\CatUsers}{6{,}000\xspace}
\newcommand{\CrActMonthsPostC}{12.32\xspace}
\newcommand{\CrActMonthsPostT}{12.37\xspace}
\newcommand{\CrAgeOlderAny}{2.7\xspace}
\newcommand{\CrAgeOlderAnyhi}{3.4\xspace}
\newcommand{\CrAgeOlderAnylo}{1.9\xspace}
\newcommand{\CrAgeOlderN}{51\xspace}
\newcommand{\CrAgeOlderRel}{6.1\xspace}
\newcommand{\CrAgeOlderRelhi}{9.3\xspace}
\newcommand{\CrAgeOlderRello}{2.9\xspace}
\newcommand{\CrAgeRecentAny}{2.9\xspace}
\newcommand{\CrAgeRecentAnyhi}{4.0\xspace}
\newcommand{\CrAgeRecentAnylo}{1.6\xspace}
\newcommand{\CrAgeRecentN}{27\xspace}
\newcommand{\CrAgeRecentRel}{6.5\xspace}
\newcommand{\CrAgeRecentRelhi}{12.0\xspace}
\newcommand{\CrAgeRecentRello}{0.7\xspace}
\newcommand{\CrAnyLong}{1.8\xspace}
\newcommand{\CrAnyLonghi}{2.6\xspace}
\newcommand{\CrAnyLonglo}{1.0\xspace}
\newcommand{\CrAnyPost}{2.8\xspace}
\newcommand{\CrAnyPostPre}{1.4\xspace}
\newcommand{\CrAnyPostPrehi}{1.9\xspace}
\newcommand{\CrAnyPostPrelo}{0.8\xspace}
\newcommand{\CrAnyPosthi}{3.4\xspace}
\newcommand{\CrAnyPostlo}{2.1\xspace}
\newcommand{\CrAnyRef}{39\xspace}
\newcommand{\CrAnyTrend}{\ensuremath{-}0.1\xspace}
\newcommand{\CrAnyTrendhi}{1.1\xspace}
\newcommand{\CrAnyTrendlo}{\ensuremath{-}1.2\xspace}
\newcommand{\CrBothRoles}{1\xspace}
\newcommand{\CrClusterCompMax}{45\xspace}
\newcommand{\CrClusterComps}{20\xspace}
\newcommand{\CrClusterMonths}{38\xspace}
\newcommand{\CrControlArtists}{301\xspace}
\newcommand{\CrControlUnits}{357\xspace}
\newcommand{\CrDiscoveryPost}{0.03\xspace}
\newcommand{\CrDiscoveryPosthi}{0.11\xspace}
\newcommand{\CrDiscoveryPostlo}{\ensuremath{-}0.05\xspace}
\newcommand{\CrEvents}{78\xspace}
\newcommand{\CrEventsAtTwentyFour}{68\xspace}
\newcommand{\CrFanMsT}{52\xspace}
\newcommand{\CrFansAllC}{606{,}175\xspace}
\newcommand{\CrFansAllT}{102{,}617\xspace}
\newcommand{\CrFansMedianT}{846\xspace}
\newcommand{\CrIntQuitA}{\ensuremath{-}2.8\xspace}
\newcommand{\CrIntQuitC}{\ensuremath{-}3.0\xspace}
\newcommand{\CrIntRelA}{17.6\xspace}
\newcommand{\CrIntRelAhi}{22.8\xspace}
\newcommand{\CrIntRelAlo}{12.2\xspace}
\newcommand{\CrIntRelC}{3.7\xspace}
\newcommand{\CrIntRelChi}{6.7\xspace}
\newcommand{\CrIntRelClo}{0.6\xspace}
\newcommand{\CrLatestPost}{\ensuremath{-}0.9\xspace}
\newcommand{\CrLatestPosthi}{0.5\xspace}
\newcommand{\CrLatestPostlo}{\ensuremath{-}2.4\xspace}
\newcommand{\CrLisGrowthC}{19\xspace}
\newcommand{\CrLisGrowthQoneT}{82\xspace}
\newcommand{\CrLisGrowthQthreeT}{190\xspace}
\newcommand{\CrLisGrowthT}{123\xspace}
\newcommand{\CrLowObsPctC}{4.0\xspace}
\newcommand{\CrLowObsPctT}{3.5\xspace}
\newcommand{\CrMainstreamPost}{\ensuremath{-}0.4\xspace}
\newcommand{\CrMainstreamPosthi}{0.0\xspace}
\newcommand{\CrMainstreamPostlo}{\ensuremath{-}0.7\xspace}
\newcommand{\CrMainstreamRef}{48.5\xspace}
\newcommand{\CrMsCutHi}{62\xspace}
\newcommand{\CrMsCutLo}{42\xspace}
\newcommand{\CrMsQuitA}{\ensuremath{-}2.5\xspace}
\newcommand{\CrMsQuitAhi}{\ensuremath{-}0.7\xspace}
\newcommand{\CrMsQuitAlo}{\ensuremath{-}4.3\xspace}
\newcommand{\CrMsQuitC}{\ensuremath{-}5.1\xspace}
\newcommand{\CrMsQuitChangeA}{\ensuremath{-}4.1\xspace}
\newcommand{\CrMsQuitChangeAhi}{\ensuremath{-}2.4\xspace}
\newcommand{\CrMsQuitChangeAlo}{\ensuremath{-}5.6\xspace}
\newcommand{\CrMsQuitChangeB}{\ensuremath{-}2.7\xspace}
\newcommand{\CrMsQuitChangeC}{\ensuremath{-}3.5\xspace}
\newcommand{\CrMsQuitChangeChi}{\ensuremath{-}1.6\xspace}
\newcommand{\CrMsQuitChangeClo}{\ensuremath{-}5.6\xspace}
\newcommand{\CrMsQuitChi}{\ensuremath{-}3.4\xspace}
\newcommand{\CrMsQuitClo}{\ensuremath{-}6.8\xspace}
\newcommand{\CrMsQuitContrast}{2.6\xspace}
\newcommand{\CrMsQuitContrastChange}{\ensuremath{-}0.6\xspace}
\newcommand{\CrMsQuitContrastChangehi}{1.7\xspace}
\newcommand{\CrMsQuitContrastChangelo}{\ensuremath{-}2.7\xspace}
\newcommand{\CrMsQuitPlaceboContrast}{3.2\xspace}
\newcommand{\CrMsQuitPlaceboContrasthi}{4.9\xspace}
\newcommand{\CrMsQuitPlaceboContrastlo}{1.4\xspace}
\newcommand{\CrMsRelContrast}{\ensuremath{-}4.0\xspace}
\newcommand{\CrMsRelContrasthi}{\ensuremath{-}0.5\xspace}
\newcommand{\CrMsRelContrastlo}{\ensuremath{-}7.7\xspace}
\newcommand{\CrMsReturnContrastChange}{0.7\xspace}
\newcommand{\CrMsReturnContrastChangehi}{3.8\xspace}
\newcommand{\CrMsReturnContrastChangelo}{\ensuremath{-}2.7\xspace}
\newcommand{\CrMsShareContrast}{0.031\xspace}
\newcommand{\CrMsShareContrasthi}{0.062\xspace}
\newcommand{\CrMsShareContrastlo}{0.002\xspace}
\newcommand{\CrNewMsSurgeT}{52\xspace}
\newcommand{\CrNewRatioC}{1.1\xspace}
\newcommand{\CrNewRatioT}{3.5\xspace}
\newcommand{\CrOlderNevents}{68\xspace}
\newcommand{\CrOlderPost}{\ensuremath{-}1.0\xspace}
\newcommand{\CrOlderPosthi}{\ensuremath{-}0.1\xspace}
\newcommand{\CrOlderPostlo}{\ensuremath{-}2.0\xspace}
\newcommand{\CrOlderRef}{45.2\xspace}
\newcommand{\CrOtherPost}{1.9\xspace}
\newcommand{\CrOtherPosthi}{3.6\xspace}
\newcommand{\CrOtherPostlo}{0.6\xspace}
\newcommand{\CrPosBaseAny}{3.1\xspace}
\newcommand{\CrPosBaseAnyhi}{3.9\xspace}
\newcommand{\CrPosBaseAnylo}{2.3\xspace}
\newcommand{\CrPosBaseShare}{\ensuremath{-}0.033\xspace}
\newcommand{\CrPosBaseSharehi}{\ensuremath{-}0.008\xspace}
\newcommand{\CrPosBaseSharelo}{\ensuremath{-}0.060\xspace}
\newcommand{\CrQuitC}{29.1\xspace}
\newcommand{\CrQuitChange}{\ensuremath{-}3.1\xspace}
\newcommand{\CrQuitChangehi}{\ensuremath{-}2.1\xspace}
\newcommand{\CrQuitChangelo}{\ensuremath{-}4.2\xspace}
\newcommand{\CrQuitDiff}{\ensuremath{-}3.3\xspace}
\newcommand{\CrQuitDiffComphi}{\ensuremath{-}2.5\xspace}
\newcommand{\CrQuitDiffComplo}{\ensuremath{-}5.0\xspace}
\newcommand{\CrQuitDiffMonthhi}{\ensuremath{-}2.3\xspace}
\newcommand{\CrQuitDiffMonthlo}{\ensuremath{-}4.4\xspace}
\newcommand{\CrQuitDiffhi}{\ensuremath{-}1.9\xspace}
\newcommand{\CrQuitDifflo}{\ensuremath{-}4.6\xspace}
\newcommand{\CrQuitJackMax}{\ensuremath{-}3.1\xspace}
\newcommand{\CrQuitJackMin}{\ensuremath{-}3.5\xspace}
\newcommand{\CrQuitPlaceboDiff}{\ensuremath{-}0.1\xspace}
\newcommand{\CrQuitPlaceboDiffhi}{1.2\xspace}
\newcommand{\CrQuitPlaceboDifflo}{\ensuremath{-}1.4\xspace}
\newcommand{\CrQuitRelPct}{11\xspace}
\newcommand{\CrQuitT}{25.8\xspace}
\newcommand{\CrRawSharePctAbsC}{55\xspace}
\newcommand{\CrRawSharePctAbsT}{51\xspace}
\newcommand{\CrRegAny}{2.8\xspace}
\newcommand{\CrRegAnyPre}{1.4\xspace}
\newcommand{\CrRegAnyPrese}{0.3\xspace}
\newcommand{\CrRegAnyse}{0.3\xspace}
\newcommand{\CrRegClusters}{378\xspace}
\newcommand{\CrRegRel}{6.5\xspace}
\newcommand{\CrRegRelPre}{3.1\xspace}
\newcommand{\CrRegRelPrese}{1.4\xspace}
\newcommand{\CrRegRelse}{1.3\xspace}
\newcommand{\CrRegShare}{\ensuremath{-}0.022\xspace}
\newcommand{\CrRegSharePre}{\ensuremath{-}0.013\xspace}
\newcommand{\CrRegSharePrese}{0.008\xspace}
\newcommand{\CrRegSharese}{0.011\xspace}
\newcommand{\CrRelCapFifty}{11.6\xspace}
\newcommand{\CrRelCapFiftyhi}{16.5\xspace}
\newcommand{\CrRelCapFiftylo}{7.0\xspace}
\newcommand{\CrRelCapFive}{4.5\xspace}
\newcommand{\CrRelCapFivehi}{6.6\xspace}
\newcommand{\CrRelCapFivelo}{2.4\xspace}
\newcommand{\CrRelCapMonthPct}{1.7\xspace}
\newcommand{\CrRelCapUnitPct}{11.9\xspace}
\newcommand{\CrRelPost}{6.2\xspace}
\newcommand{\CrRelPostPre}{3.5\xspace}
\newcommand{\CrRelPostPrehi}{5.9\xspace}
\newcommand{\CrRelPostPrelo}{1.3\xspace}
\newcommand{\CrRelPosthi}{9.1\xspace}
\newcommand{\CrRelPostlo}{3.3\xspace}
\newcommand{\CrRelTrend}{\ensuremath{-}1.0\xspace}
\newcommand{\CrRelTrendhi}{5.6\xspace}
\newcommand{\CrRelTrendlo}{\ensuremath{-}7.1\xspace}
\newcommand{\CrRetDiD}{0.3\xspace}
\newcommand{\CrRetDiDhi}{2.5\xspace}
\newcommand{\CrRetDiDlo}{\ensuremath{-}2.2\xspace}
\newcommand{\CrRetMinRef}{20\xspace}
\newcommand{\CrRetNearDiD}{\ensuremath{-}1.6\xspace}
\newcommand{\CrRetNearDiDhi}{1.0\xspace}
\newcommand{\CrRetNearDiDlo}{\ensuremath{-}4.1\xspace}
\newcommand{\CrRetNrefT}{105{,}420\xspace}
\newcommand{\CrRetNsurgeT}{176{,}467\xspace}
\newcommand{\CrRetRefC}{39\xspace}
\newcommand{\CrRetRefT}{44\xspace}
\newcommand{\CrRetSurgeC}{34\xspace}
\newcommand{\CrRetSurgeT}{39\xspace}
\newcommand{\CrReturnC}{25.0\xspace}
\newcommand{\CrReturnChange}{5.1\xspace}
\newcommand{\CrReturnChangehi}{7.0\xspace}
\newcommand{\CrReturnChangelo}{3.4\xspace}
\newcommand{\CrReturnDiff}{4.6\xspace}
\newcommand{\CrReturnDiffhi}{6.3\xspace}
\newcommand{\CrReturnDifflo}{3.0\xspace}
\newcommand{\CrReturnPlaceboDiff}{\ensuremath{-}0.5\xspace}
\newcommand{\CrReturnPlaceboDiffhi}{0.9\xspace}
\newcommand{\CrReturnPlaceboDifflo}{\ensuremath{-}1.9\xspace}
\newcommand{\CrReturnT}{29.6\xspace}
\newcommand{\CrShareLong}{\ensuremath{-}0.032\xspace}
\newcommand{\CrShareLonghi}{\ensuremath{-}0.001\xspace}
\newcommand{\CrShareLonglo}{\ensuremath{-}0.065\xspace}
\newcommand{\CrSharePct}{\ensuremath{-}5.5\xspace}
\newcommand{\CrSharePcthi}{\ensuremath{-}0.4\xspace}
\newcommand{\CrSharePctlo}{\ensuremath{-}10.8\xspace}
\newcommand{\CrSharePost}{\ensuremath{-}0.024\xspace}
\newcommand{\CrSharePostPre}{\ensuremath{-}0.010\xspace}
\newcommand{\CrSharePostPrehi}{0.005\xspace}
\newcommand{\CrSharePostPrelo}{\ensuremath{-}0.028\xspace}
\newcommand{\CrSharePre}{\ensuremath{-}0.014\xspace}
\newcommand{\CrSharePrehi}{0.002\xspace}
\newcommand{\CrSharePrelo}{\ensuremath{-}0.030\xspace}
\newcommand{\CrShareRef}{0.44\xspace}
\newcommand{\CrShareTrend}{0.031\xspace}
\newcommand{\CrShareTrendhi}{0.064\xspace}
\newcommand{\CrShareTrendlo}{\ensuremath{-}0.002\xspace}
\newcommand{\CrStrictAny}{2.2\xspace}
\newcommand{\CrStrictAnyhi}{3.2\xspace}
\newcommand{\CrStrictAnylo}{1.3\xspace}
\newcommand{\CrStrictPctT}{32\xspace}
\newcommand{\CrStrictQuitDiff}{\ensuremath{-}3.9\xspace}
\newcommand{\CrStrictQuitDiffhi}{\ensuremath{-}2.6\xspace}
\newcommand{\CrStrictQuitDifflo}{\ensuremath{-}5.4\xspace}
\newcommand{\CrStrictRel}{8.4\xspace}
\newcommand{\CrStrictRelhi}{12.4\xspace}
\newcommand{\CrStrictRello}{4.3\xspace}
\newcommand{\CrTenCutHi}{25\xspace}
\newcommand{\CrTenCutLo}{17\xspace}
\newcommand{\CrTenQuitChangeA}{\ensuremath{-}3.4\xspace}
\newcommand{\CrTenQuitChangeC}{\ensuremath{-}3.4\xspace}
\newcommand{\CrTenQuitContrastChange}{0.1\xspace}
\newcommand{\CrTenQuitContrastChangehi}{1.9\xspace}
\newcommand{\CrTenQuitContrastChangelo}{\ensuremath{-}1.7\xspace}
\newcommand{\CrTenReturnChangeA}{7.7\xspace}
\newcommand{\CrTenReturnChangeC}{3.1\xspace}
\newcommand{\CrTenReturnContrastChange}{5.2\xspace}
\newcommand{\CrTenReturnContrastChangehi}{8.7\xspace}
\newcommand{\CrTenReturnContrastChangelo}{1.9\xspace}
\newcommand{\CrTenShareA}{\ensuremath{-}0.028\xspace}
\newcommand{\CrTenShareAhi}{0.003\xspace}
\newcommand{\CrTenShareAlo}{\ensuremath{-}0.056\xspace}
\newcommand{\CrTenShareB}{0.001\xspace}
\newcommand{\CrTenShareBhi}{0.022\xspace}
\newcommand{\CrTenShareBlo}{\ensuremath{-}0.022\xspace}
\newcommand{\CrTenShareC}{\ensuremath{-}0.014\xspace}
\newcommand{\CrTenShareChi}{0.006\xspace}
\newcommand{\CrTenShareClo}{\ensuremath{-}0.034\xspace}
\newcommand{\CrTenSharePostPreA}{\ensuremath{-}0.005\xspace}
\newcommand{\CrTenSharePostPreB}{\ensuremath{-}0.009\xspace}
\newcommand{\CrTenSharePostPreC}{\ensuremath{-}0.015\xspace}
\newcommand{\CrTenShareRefA}{0.61\xspace}
\newcommand{\CrUsersDistinct}{228{,}620\xspace}
\newcommand{\CrWithinAny}{1.1\xspace}
\newcommand{\CrWithinAnyhi}{2.9\xspace}
\newcommand{\CrWithinAnylo}{\ensuremath{-}0.5\xspace}
\newcommand{\CrWithinFans}{17{,}794\xspace}
\newcommand{\CrWithinRel}{5.4\xspace}
\newcommand{\CrWithinRelhi}{13.7\xspace}
\newcommand{\CrWithinRello}{\ensuremath{-}3.2\xspace}
\newcommand{\CrWithinShare}{\ensuremath{-}0.019\xspace}
\newcommand{\CrWithinSharehi}{0.022\xspace}
\newcommand{\CrWithinSharelo}{\ensuremath{-}0.069\xspace}
\newcommand{\CrWithinTopPct}{63\xspace}
\newcommand{\CrZeroBasePctC}{19\xspace}
\newcommand{\CrZeroBasePctT}{19\xspace}
\newcommand{\LedgerDocumented}{6\xspace}
\newcommand{\MechDisCutHi}{4.8\xspace}
\newcommand{\MechDisMedian}{1.6\xspace}
\newcommand{\MechDisPosN}{46\xspace}
\newcommand{\MechDisQone}{\ensuremath{-}1.6\xspace}
\newcommand{\MechDisQthree}{6.4\xspace}
\newcommand{\MechDisQuitChgDiff}{\ensuremath{-}0.2\xspace}
\newcommand{\MechDisQuitChgDiffhi}{2.2\xspace}
\newcommand{\MechDisQuitChgDifflo}{\ensuremath{-}2.4\xspace}
\newcommand{\MechDisQuitChgHi}{\ensuremath{-}2.7\xspace}
\newcommand{\MechDisQuitChgHihi}{\ensuremath{-}0.9\xspace}
\newcommand{\MechDisQuitChgHilo}{\ensuremath{-}4.6\xspace}
\newcommand{\MechDisQuitChgLo}{\ensuremath{-}2.6\xspace}
\newcommand{\MechDisQuitChgLohi}{\ensuremath{-}1.1\xspace}
\newcommand{\MechDisQuitChgLolo}{\ensuremath{-}4.2\xspace}
\newcommand{\MechDisQuitHi}{\ensuremath{-}1.6\xspace}
\newcommand{\MechDisQuitLo}{\ensuremath{-}4.6\xspace}
\newcommand{\MechDisRetChgDiff}{0.6\xspace}
\newcommand{\MechDisRetChgDiffhi}{4.6\xspace}
\newcommand{\MechDisRetChgDifflo}{\ensuremath{-}3.1\xspace}
\newcommand{\MechEvents}{78\xspace}
\newcommand{\MechGrowRetChgDiff}{4.9\xspace}
\newcommand{\MechGrowRetChgDiffhi}{9.3\xspace}
\newcommand{\MechGrowRetChgDifflo}{0.7\xspace}
\newcommand{\MechNrCutHi}{4.0\xspace}
\newcommand{\MechNrCutLo}{3.3\xspace}
\newcommand{\MechNrQuitChgDiff}{\ensuremath{-}1.3\xspace}
\newcommand{\MechNrQuitChgDiffhi}{1.4\xspace}
\newcommand{\MechNrQuitChgDifflo}{\ensuremath{-}4.1\xspace}
\newcommand{\MechNrRetChgDiff}{6.2\xspace}
\newcommand{\MechNrRetChgDiffhi}{11.2\xspace}
\newcommand{\MechNrRetChgDifflo}{1.6\xspace}
\newcommand{\MechNrRetChgHi}{8.4\xspace}
\newcommand{\MechNrRetChgHihi}{12.1\xspace}
\newcommand{\MechNrRetChgHilo}{5.1\xspace}
\newcommand{\MechNrRetChgLo}{2.2\xspace}
\newcommand{\MechNrRetChgLohi}{5.3\xspace}
\newcommand{\MechNrRetChgLolo}{\ensuremath{-}0.9\xspace}
\newcommand{\MechNrRetChgMid}{4.6\xspace}
\newcommand{\MechPplPctDiffAbs}{9\xspace}
\newcommand{\MechPplPctDiffhiAbs}{3\xspace}
\newcommand{\MechPplPctDiffloAbs}{13\xspace}
\newcommand{\MechRecCueKnownPctC}{83\xspace}
\newcommand{\MechRecCueKnownPctT}{61\xspace}
\newcommand{\MechRecCueLate}{2.1\xspace}
\newcommand{\MechRecCueLatehi}{3.0\xspace}
\newcommand{\MechRecCueLatelo}{1.3\xspace}
\newcommand{\MechRecCueNotTopN}{49\xspace}
\newcommand{\MechRecCuePost}{3.0\xspace}
\newcommand{\MechRecCuePosthi}{4.0\xspace}
\newcommand{\MechRecCuePostlo}{2.1\xspace}
\newcommand{\MechRecCuePre}{0.2\xspace}
\newcommand{\MechRecCuePrehi}{0.5\xspace}
\newcommand{\MechRecCuePrelo}{\ensuremath{-}0.1\xspace}
\newcommand{\MechRecCueRawC}{0.7\xspace}
\newcommand{\MechRecCueRawT}{3.7\xspace}
\newcommand{\MechRecCueRefT}{8.4\xspace}
\newcommand{\MechRecCueTopOnly}{2.1\xspace}
\newcommand{\MechRecCueTopOnlyN}{29\xspace}
\newcommand{\MechRecCueTopOnlyhi}{3.5\xspace}
\newcommand{\MechRecCueTopOnlylo}{0.8\xspace}
\newcommand{\MechRecDeepPost}{0.3\xspace}
\newcommand{\MechRecDeepPosthi}{1.2\xspace}
\newcommand{\MechRecDeepPostlo}{\ensuremath{-}0.6\xspace}
\newcommand{\MechRecDeepRefT}{39.6\xspace}
\newcommand{\MechRecEvents}{78\xspace}
\newcommand{\OfficialChanged}{3\xspace}
\newcommand{\PastAuditQuit}{\ensuremath{-}4.2\xspace}
\newcommand{\PastAuditQuithi}{\ensuremath{-}2.4\xspace}
\newcommand{\PastAuditQuitlo}{\ensuremath{-}5.8\xspace}
\newcommand{\PastCtrlAlbPct}{24\xspace}
\newcommand{\PastCtrlSurgePct}{45\xspace}
\newcommand{\PastSameEvents}{78\xspace}
\newcommand{\PastSameQuit}{\ensuremath{-}2.9\xspace}
\newcommand{\PastSameQuithi}{\ensuremath{-}1.7\xspace}
\newcommand{\PastSameQuitlo}{\ensuremath{-}4.1\xspace}
\newcommand{\RobQuitMax}{\ensuremath{-}2.3\xspace}
\newcommand{\RobQuitMin}{\ensuremath{-}5.7\xspace}
\newcommand{\RobVariants}{13\xspace}
\newcommand{\minLis}{300\xspace}
\newcommand{\nPlaysAllB}{21.1\xspace}
\newcommand{\nShards}{16\xspace}
\newcommand{\nUniverse}{10{,}022\xspace}
\newcommand{\nUsersActive}{594{,}415\xspace}
\newcommand{\nUsersAll}{594{,}415\xspace}



\title{Before They Were Cool: Do Early Fans Abandon Artists Who Suddenly Become Popular? Evidence from Online Music Listening}

\author{
  Zihan Chen \\
  Stevens Institute of Technology \\
  Hoboken, NJ, USA \\
  \texttt{zchen61@stevens.edu} \\
  \And
  Lei Nico Zheng \\
  University of Massachusetts Boston \\
  Boston, MA, USA \\
  \texttt{lei.zheng@umb.edu} \\
  \And
  Di Zhu \\
  Stevens Institute of Technology \\
  Hoboken, NJ, USA \\
  \texttt{dzhu1@stevens.edu} \\
}

\begin{document}
\maketitle

\begin{abstract}
Identity-signalling theory, echoed in the hipster's ``I liked them before they were cool'', expects an artist's first fans to pull away once the artist becomes popular, by leaving or by avoiding what the newcomers play. A rival account expects the event that brings the crowd to renew existing listeners' attention to the artist. We test both in the Last.fm listening histories of \nUsersAll users (\nPlaysAllB billion plays, 2005--2013), where every play appears on the user's profile. Study~1 detects \CrEvents crowd arrivals: months in which the number of people discovering an artist tripled without a new album, single or EP, for example after a song appeared in a film. Study~2 compares Grammy and Mercury Prize winners with nominees who won nothing. We compare listeners who had played the artist before the crowd arrived (early fans) with early fans of comparison artists. Early fans did not withdraw. Those still listening stopped less often (\CrQuitT\% vs.\ \CrQuitC\% over six months), and those who had drifted away returned more often (\CrReturnT\% vs.\ \CrReturnC\%). Those who kept listening converged on the crowd: the song most newcomers played took a larger share of their plays, with no turn to deep cuts. More lapsed fans returned after larger surges, and neither fans with the most distinctive tastes nor fans facing a more mainstream crowd quit more. Award wins brought few new listeners and no clear change. Where listening is displayed by default, early fans moved towards the crowd after sudden popularity, not away from it.
\end{abstract}


\section{Introduction}
\label{sec:intro}

Within hours of Arcade Fire winning Album of the Year at the 2011 Grammy Awards, a Tumblr blog called \emph{Who Is Arcade Fire??!!?} was cataloguing the indignant posts of Twitter and Facebook users who had never heard of the band \citep{Koski2011}. For people who had followed the band since its first album, the night meant something else: a group they had loved as a discovery now belonged to everyone. Popular culture has a stock phrase for what such fans are supposed to do next. \emph{I liked them before they were cool} implies that the early fan moves on once the crowd arrives.

The phrase has serious theory behind it. Music preferences are among the clearest signals of identity that people send \citep{Rentfrow2003,Rentfrow2006,Berger2007}. Identity-signalling theory predicts that people abandon a taste once others adopt it, especially others they do not want to be mistaken for, so that the taste keeps saying something about who they are \citep{Berger2008}. The prediction is conditional on visibility: its authors expect people to diverge in public choices while perhaps continuing ``to listen secretly in private'' \citep[p.~598]{Berger2008}. Online platforms have blurred that line. Research in this journal has shown that identity motives decide which music people choose to share \citep{Johnson2018}. Last.fm records every song a user plays and shows it on the user's profile, and its users change what they listen to so as to control the impression their profiles give \citep{Silfverberg2011}. A play on such a platform is chosen for oneself but leaves a public trace by default. Whether identity concerns reach behaviour of this kind is open.

The evidence for abandonment comes from visible choices such as wristbands \citep{Berger2008}, from the rise and fall of first names \citep{Berger2009}, and from ratings: books that win major prizes are rated lower afterwards than books on the same shortlists \citep{Kovacs2014}. Ratings come from whoever chooses to rate, and a prize brings new readers with different tastes, so a fall in average ratings can reflect a change in the audience without any change in how the original audience behaves. We know of no study that follows the listening of an artist's existing audience through a sudden rise in popularity.

A second account predicts the opposite. The film scene, television episode or viral clip that brings newcomers to an artist also reaches the people who already know it, and the artist then turns up in charts, recommendations and conversations they see too. Such events raise demand beyond first-time listeners: concerts raise listening among attendees and their friends \citep{Ternovski2020}, entry into a streaming chart raises repeat listening \citep{Sim2022}, and songs reused in films gain listeners on streaming services \citep{Aguiar2024}. On this renewed-attention account, early fans should move with the crowd. The two accounts disagree on two questions that listening records can answer. Do early fans pull away from the artist, by stopping or by failing to come back? And do the fans who keep listening pull away from the crowd, by avoiding the song the newcomers came for in favour of the earlier records and the deep cuts? The identity account answers yes to both and expects the response to be strongest among the fans with the most distinctive tastes, where the crowd is more mainstream than they are and where the surge is largest. The renewed-attention account answers no, expects returning among fans who had drifted away to grow with the size of the surge, and expects fans to play the song that drew the crowd.

We test the two accounts with the Last.fm listening histories of \nUsersAll users, \nPlaysAllB billion time-stamped plays from 2005 to 2013 that are linked to the MusicBrainz database of artists, releases and recordings. We study two kinds of events. In Study~1 we detect \emph{crowd arrivals}: months in which the number of people discovering an artist tripled while the artist released no new album, single or EP, for example because a song reached a wide audience through a film or a television series. Across the \CrEvents arrivals we analyse, newcomers in the first three months were \CrNewRatioT times their usual number and monthly listeners grew by a median of \CrLisGrowthT\%, against \CrLisGrowthC\% for matched comparison artists. Most of these artists were established: for \CrAgeOlderN of them the latest studio album was more than two years old, so the crowd arrived for records the early fans already knew. In Study~2 we compare winners of the Grammy Awards and the Mercury Prize with the nominees who lost, a status shock whose effect on the size of the audience is itself an empirical question. In both studies we follow the artist's \emph{early fans}, people who had played the artist repeatedly before the crowd arrived, for up to two years afterwards. We compare them with the early fans of comparison artists: in Study~1, artists at the same point of their release cycle; in Study~2, the nominees who lost.

Early fans moved with the crowd. Those who were still listening before the event were less likely than comparison fans to stop: \CrQuitT\% vs.\ \CrQuitC\% did not play the artist at all in the following six months, a rate \CrQuitRelPct\% lower. Early fans who had drifted away were more likely to come back (\CrReturnT\% vs.\ \CrReturnC\%). Neither gap existed in the six months before the event, and the result holds in all \RobVariants alternative specifications we report, including comparison artists chosen only with information available before the event and the removal of every artist with any new recording near the event.

Early fans who kept listening converged on the crowd: the share of their plays going to the song most newcomers played rose by \MechRecCuePost percentage points relative to comparison fans, from \MechRecCueRefT\%. The share going to the artist's less-played recordings did not change. The fans with the most distinctive tastes changed no differently from the others, and quitting did not rise where the newcomers were more mainstream than the early fans. Returning grew with the size of the surge: it rose by \MechNrRetChgDiff points more in the third of events with the largest surge of newcomers than in the third with the smallest. How much early fans listened changed little beyond trends that had begun before the event, and the gains came mostly from light listeners and from fans who had found the artist recently. The newcomers who arrived in a surge kept listening about as often as those who had found the artist before, while the artist's plays per listener, counted over its whole audience, fell by \MechPplPctDiffAbs\% relative to comparison artists. Award wins brought winners few additional listeners on Last.fm and no clear change in their early fans.

The paper makes three contributions. First, it constrains how far identity-signalling theory extends to ordinary digital behaviour. When an artist's audience grew suddenly, the people who already listened to it did not withdraw from it or from the crowd's song, on a platform that displays every play and whose users are known to manage their plays for their profiles. A large influx of adopters was not enough to produce behavioural abandonment, so the theory's abandonment prediction should be scoped to deliberate displays and to choices made in view of a known, dissimilar group. Second, it shows what took the place of withdrawal. The pattern is one of renewed attention in the existing audience: lapsed fans returned, more of them the larger the surge, and fans who kept listening converged on the crowd's song. Research on concerts, charts and film placements has estimated how such events raise demand \citep{Ternovski2020,Sim2022,Aguiar2024}; following identifiable prior listeners shows who in the existing audience responds, whether lapsed or active, and to which recording. Third, it shows how averages mislead about the first audience. A per-listener average fell after the crowd arrived while the first listeners quit less, and a comparison without matched artists would have read the ordinary decay of listening after a release as an exodus. Following the same people over time, and comparing artists at the same point of their release cycle, separates who is listening from how they listen.

\section{Theory and competing predictions}
\label{sec:background}

\subsection{Taste as a displayed identity}

Few consumption choices say as much about a person as the music they like. Music preferences are tied to personality and to self-views such as political orientation \citep{Rentfrow2003}; observers form accurate impressions of strangers from their music preferences, and music is the most common topic of conversation when strangers get acquainted \citep{Rentfrow2006}. Taste also draws boundaries: the genres people reject map onto social distinctions as clearly as the genres they embrace \citep{Bryson1996,Bourdieu1984}.

Online platforms turned these preferences into displays. Profile pages make lists of favourite artists into ``taste performances'' that convey prestige, differentiation, authenticity or a theatrical persona \citep{Liu2007}, and research in this journal has shown that identity motives decide which music people show. In an experiment, people motivated to fit in with a group's tastes said they would share less unique music and film, people motivated to present their best selves chose more prestigious titles, and individual differences in the need for uniqueness moderated some of these effects \citep{Johnson2018}. Displaying a taste can in turn change it: public sharing strengthened the attitudes of people who often display their tastes online and weakened those of people who rarely do \citep{Johnson2015}. Whose view matters is specific. Sharing a message online made people more committed to it only through the belief that their close ties now knew about it \citep{Lew2024}.

Last.fm, the service whose logs we analyse, makes listening itself a display. Every play is recorded (``scrobbled'') and shown on the user's profile, which lists the most recently played songs and regularly updated charts of the user's top artists; the site also suggests musically similar users, called neighbours, from overlapping listening \citep{Baym2009}. The display is automatic, but it is not ignored. In an exploratory study of Last.fm users, people changed what they listened to, not only what they published, to control the self-presentation their profiles produced \citep{Silfverberg2011}. Automated and manual sharing are judged by different norms, and the common goal of such profile work is to appear authentic \citep{Uski2016}. Users of today's streaming services describe the same awareness once listening that used to stay at home becomes visible to others \citep{Walsh2024}. On Last.fm, a play is chosen for oneself but leaves a persistent trace that friends and neighbours can read. A user who has played an artist for years carries that history on their profile, and when the artist becomes famous, what the history says about the user changes without the user doing anything.

\subsection{The identity account: why early fans might pull away}

Identity-signalling theory holds that people use tastes to communicate who they are, so a taste that spreads to others loses its value as a signal. People diverge from majorities more in domains seen as symbolic of identity, such as music, than in functional ones, and in such domains they abandon preferences once majorities come to share them \citep{Berger2007}. They abandon a taste most readily when it is adopted by a group they do not want to be mistaken for \citep{Berger2008}. The theory gives a functional reason for older ideas: the snob effect, in which demand falls as others' demand rises \citep{Leibenstein1950}, the need for uniqueness \citep{Snyder1977,Tian2001} and the wish to belong to groups that stay distinct \citep{Brewer1991}.

The theory's prediction is conditional on what others can see. \citet[p.~598]{Berger2008} note that people who would not want others to see them listening to an artist adopted by a dissimilar group might, if they really liked the artist, ``continue to listen secretly in private''; divergence should appear in public choices more than in private ones. A Last.fm play sits between the two cases. It is chosen for private enjoyment, yet it is displayed by default to the people who look at the user's profile, and Last.fm users are known to adjust their plays for that audience \citep{Silfverberg2011}. If identity concerns reach behaviour that leaves a public trace, they should reach it here.

Field evidence supports the abandonment prediction. First names that rose quickly in popularity also fell quickly, and people view fast-rising cultural items as fads \citep{Berger2009}. Goodreads ratings of books that won major prizes fell after the award relative to the ratings of books on the same shortlists; the authors trace the fall to a larger and more diverse audience and to an increase in popularity that is ``off-putting to some'' \citep{Kovacs2014}. Indie consumers guard their identity investments against devaluation by the mainstream ``hipster'' label \citep{Arsel2011}, and listeners with cultural capital at stake experience streaming personalisation, which does their curating for them, as a threat to distinction \citep{Webster2023}.

The account offers early fans two ways to keep their distance. They can leave: when an artist's audience suddenly grows, early fans should become more likely to stop playing it, and fans who had drifted away less likely to return. The response should be strongest among fans whose tastes are the most distinctive and where the newcomers are a group they do not want to be mistaken for \citep{Berger2008}. Or they can stay distinct without leaving. Insiders prefer signals that outsiders do not recognise \citep{Berger2010}, people conform on one dimension while differentiating on another \citep{Chan2012}, and indie consumers protect their investments through practices other than exit \citep{Arsel2011}. Early fans who keep listening should then turn away from the song the newcomers came for, towards the artist's earlier records and less-played recordings. Identity motives can also favour staying, since the claim to have been there first requires having been a fan. That claim lives in conversation and in deliberate displays, which listening records do not show, so we test the two routes that do appear in them.

\subsection{The renewed-attention account: why early fans might move with the crowd}

The event that brings newcomers also reaches the people who already know the artist. A film scene, a television episode or a viral clip plays the song to existing listeners as well as new ones, and the artist then appears in charts, recommendations, radio and conversation. Each of these brings the artist back to the attention of people for whom it had faded. Such events raise demand among people who are not newcomers too. On Last.fm, attending an artist's concert raised attendees' listening to the artist, and the increase spread to their friends for popular artists \citep{Ternovski2020}. On a streaming platform, a song's entry into the hourly top-100 chart increased its discovery and its repeat consumption, a pattern the authors attribute to salience more than to learning from others' choices \citep{Sim2022}. Songs reused in films were more often discovered when the films drew larger audiences, and the reuse raised their consumption on streaming charts \citep{Aguiar2024}. Recency is the strongest predictor of whether people consume an item again \citep{Anderson2014}, and where listeners see the growing audience, its size can add to the pull, as view counts do for the enjoyment of digital media \citep{Waddell2020}.

We call the interpretation that covers these routes \emph{renewed attention}: the event makes the artist present again for people who already know it, whether they hear the song, see the artist in charts or notice that others now listen. On this account early fans should be less likely to stop playing the artist and, above all, those who had drifted away should be more likely to return. The change should be largest where the artist had left a fan's regular rotation, among lapsed fans, light listeners and fans whose tie to the artist is recent, and smallest among heavy listeners who were playing the artist every month anyway. It does not depend on how distinctive a fan's taste is. A larger surge of newcomers marks an event that reached more people, so returning should grow with it. Early fans who play the artist should turn towards the recording that drew the crowd, since that is what the event put back in front of them.

The two accounts are not exclusive. Popularity can dilute a signal and widen an audience at the same time; on a fashion platform, for example, fast-fashion copycats both cannibalised premium brands and expanded their presence in users' posts \citep{Shi2023}. The question is which dominates for the people who were there first, and the pattern of who responds, to what and by how much tells the accounts apart.

\subsection{Other accounts}

Two further accounts make distinct predictions, which our data test through behavioural proxies. If early fans tire of an artist because they hear it everywhere, quitting should rise, plausibly most among fans with mainstream tastes, to the extent that they are the most exposed to radio, charts and advertising. If fans bask in the reflected glory of an artist's success \citep{Cialdini1976}, staying should rise most among long-standing fans, for whom having been there first is worth most.

\subsection{Competing predictions}

Table~\ref{tab:predictions} sets out what each account predicts for the behaviours we observe. We state the predictions of the two main accounts as competing hypotheses. H1 and H4 carry the comparison: whether early fans pull away from the artist, and whether those who stay pull away from the crowd's song.

\begin{quote}
\textbf{H1 (withdrawal or return).} When an artist's audience suddenly grows, its early fans become more likely to stop playing it and lapsed fans less likely to return (H1a, identity account), or less likely to stop and more likely to return (H1b, renewed-attention account).

\textbf{H2 (who).} The change is concentrated among early fans with the most distinctive tastes (H2a), or among early fans whose listening had lapsed or thinned, whatever their taste (H2b).

\textbf{H3 (the crowd).} Quitting rises more where the surge of newcomers is larger and where the newcomers' tastes are more mainstream than the early fans' (H3a), or returning rises more where the surge is larger (H3b).

\textbf{H4 (which recordings).} Early fans who keep listening turn away from the recording that drew the crowd, towards the artist's earlier records and less-played recordings (H4a), or towards the recording that drew the crowd (H4b).
\end{quote}

We also ask whether listeners who discover an artist in a surge leave it sooner than earlier discoverers. Names that rose quickly in popularity fell quickly \citep{Berger2009}, which concerns successive cohorts of adopters; whether the same adopters lapse faster is a separate question about the crowd.

\begin{table}[t]
\centering
\small
\caption{What four accounts predict for early fans when an artist's audience suddenly grows.}
\label{tab:predictions}
\begin{tabularx}{\textwidth}{>{\raggedright\arraybackslash}p{0.2\textwidth}>{\raggedright\arraybackslash}X>{\raggedright\arraybackslash}X>{\raggedright\arraybackslash}X>{\raggedright\arraybackslash}X}
\toprule
Observable & Identity signalling & Renewed attention & Satiation & Reflected glory \\
\midrule
Quitting; returning among lapsed fans (H1) & Rises; falls & Falls; rises & Rises; no prediction & Falls; rises \\
Who changes most (H2) & Fans with the most distinctive tastes & Lapsed, light and recent fans, whatever their taste & Fans with mainstream tastes & Long-standing fans \\
Larger surge; crowd more mainstream than the fans (H3) & More quitting; more quitting & More returning; no prediction & More quitting; no prediction & No prediction \\
Which recordings (H4) & Away from the recording that drew the crowd, towards earlier and less-played ones & Towards the recording that drew the crowd & Away from the recording everyone hears & No prediction \\
\bottomrule
\end{tabularx}
\end{table}

We measure the distinctiveness of a fan's taste by how far their listening lies from the most popular artists, a behavioural measure rather than a measure of the need for uniqueness itself. We measure the makeup of the crowd by comparing the newcomers' listening with the early fans' on the same measure. The order of the work matters for reading these tests. The analysis plan, written before the full data were processed, framed the tests around the identity account. It specified the tests of taste distinctiveness, tenure and listening intensity, the size of the surge, the mainstream gap between newcomers and early fans, the album split and the newcomers' retention. Quitting and returning became the primary endpoint once the first results were in, and the recording-level test of H4 was added after the main results were known.

\section{Data and design}
\label{sec:data}

\subsection{Listening histories}

We use MLHD+, the MetaBrainz Foundation's cleaned release of the Music Listening Histories Dataset \citep{Vigliensoni2017,MLHDplus2023}. It holds the Last.fm listening histories of \nUsersAll users, identified only by random identifiers. We use the dataset's ``complete'' files, which contain the plays whose recording MLHD+ matched to a canonical MusicBrainz recording, each with identifiers for the artist, the release and the recording, so that the same artist is recognised however its name was spelled. Plays without such a match sit in separate ``partial'' files, which we do not use. Between 2005 and 2013 these files contain \nPlaysAllB billion time-stamped plays. The histories were collected around 2013: the number of active users grows steadily until 2012 and falls in 2013 as collection ends. All events in this paper therefore fall between January 2007 and June 2012, the first year of follow-up ends by mid-2013, and the second year is observed in full only for earlier events.

We read every user file and keep three summaries. For each user and month we record the total number of plays, the plays of that year's 1,000 most-listened artists, the number of distinct artists and the number of artists played for the first time. For each user, artist and month we record the plays of the artist, for a universe of \nUniverse artists (the most-listened artists plus every award nominee; Appendix~\ref{app:data}) and for user--artist pairs with at least ten plays over the user's history. These plays are split by the release that MLHD+ links each play to: the artist's latest studio album, its earlier studio albums, or any other release (Appendix~\ref{app:data}). For each artist and month we count listeners, plays and newcomers: users who played the artist for the first time and had at least six months of listening history before, so that people who simply joined Last.fm are not counted as discovering an artist. For newcomers we also record whether they played the artist again three to twelve months later.

Release information comes from the MusicBrainz database \citep{MusicBrainz2026}. It gives each release group (an album, single or EP with all its editions) a type and a first release date, links every recording to the releases it appears on, and gives each artist an end date when a person has died or a group has disbanded. About a quarter of release-group dates are recorded to the year only; we treat such a release as possibly falling in any month of that year whenever a rule requires the absence of a new release.

\subsection{Early fans and outcomes}

For an artist and an event month $T$, the \emph{early fans} are users who played the artist at least ten times in months $T-24$ to $T-7$ and who used Last.fm in at least nine of the twelve months before the event. They are, in short, people who had been listening to the artist repeatedly before its audience grew. Because a crowd can arrive years after an artist's first records, early fans are those who were there before the crowd, not necessarily since the beginning. The gap between the selection window and the event keeps the six months just before the event free of the selection, so they provide a baseline and a test of pre-trends. A stricter definition, used as a check, requires that a fan first played the artist at least 24 months before the event and in at least three separate months of the selection window.

We follow each early fan from twelve months before to twenty-four months after the event. The \emph{primary endpoint} is quitting: among early fans who played the artist in the six months before the event (months $-6$ to $-1$), we record whether they did not play it at all in the six months after (months $0$ to $5$). The mirror outcome is returning: among early fans who had not played the artist in months $-6$ to $-1$ (lapsed fans), whether they played it in months $0$ to $5$. Both outcomes require at least four months of Last.fm use in each six-month window, so that a fan who stopped using the service is not counted as having stopped listening to the artist. The same transitions one window earlier, from months $-12..{-7}$ to $-6..{-1}$, are the placebos.

Three secondary outcomes describe how much early fans listened, measured in every month in which a fan used Last.fm at all:
\begin{itemize}
  \item \emph{listened}: whether the fan played the artist at all that month;
  \item \emph{share}: the artist's share of the fan's plays, in percent;
  \item \emph{relative share}: the share divided by the fan's own average share in months $-12$ to $-7$ (capped at ten times that average). It measures each fan's listening against their own baseline, so that light and heavy listeners count equally, and it is defined only for fans who played the artist in those months.
\end{itemize}
We describe each early fan's general taste by their \emph{mainstream share}: the share of their plays of \emph{other} artists in the selection window that went to the year's 1,000 most-listened artists. Low values mark fans whose listening lies away from what most users play.

\subsection{Study 1: crowd arrivals}

The first design finds, in the data, months in which an artist's audience suddenly grew while the artist released nothing new. An artist has a \emph{crowd arrival} in month $T$ (between January 2007 and June 2012) when four conditions hold:
\begin{enumerate}
  \item \emph{Surge.} The average number of newcomers per month in months $T$ to $T+2$ is at least three times the monthly average of the previous twelve months, and month $T$ alone is at least twice that average.
  \item \emph{An existing audience.} The artist averaged at least \minLis monthly listeners in months $T-12$ to $T-7$.
  \item \emph{No new release.} MusicBrainz lists no release group of any type credited to the artist in months $T-3$ to $T+1$ and no studio album in months $T-6$ to $T+12$.
  \item \emph{No other reason.} The artist did not die or disband in months $T-24$ to $T+12$, and had no comparable surge a year before or a year after, which removes seasonal music such as Christmas songs.
\end{enumerate}
We keep the largest qualifying surge for each artist. These rules leave the reason for the surge open: a song placed in a film or a television series, a performance, an award, a viral moment. What they rule out is the most common reason for a jump in listeners, a new release, which would also change how existing fans listen. Because release groups are credited to their main artist, the rules could miss a new recording that appeared on another artist's release, such as a soundtrack. We therefore also audit every artist-event (an artist at a given event month), treated and comparison, at the level of recordings, flagging those with a recording under any credit that first appeared in months $T-3$ to $T+1$, and report results without the flagged artist-events.

Each artist with a crowd arrival (the \emph{treated} artist) is compared with up to five \emph{comparison artists} in the same calendar month. Comparison artists had no surge (no month with even 1.5 times their usual newcomers) from a year before to a year after $T$, an existing audience, and no studio album in months $T-6$ to $T+12$. They are matched on the time since their latest studio album, within three months, so that treated and comparison artists are at the same point in the rise and decay of listening that follows a release, and among those we take the five closest in audience size. The requirements of no surge and no album after $T$ use information from after the event, so we also report results with comparison artists chosen only on information up to month $T-1$.

\subsection{Study 2: award wins}

The second design uses the timing of award ceremonies. Awards shift attention to the winner on a known night, and nominees who lose are a natural comparison: they were judged worthy at the same time. We collect all nominees for three Grammy categories (Album of the Year, Best New Artist and Best Alternative Music Album) at the ceremonies of 2008 to 2012, and the twelve-album shortlists of the Mercury Prize for 2007 to 2011, from Wikipedia. Within each ceremony, winners of these categories are compared with the nominees who won nothing that night; at the Grammys, this excludes nominees who won in any other category. The event month is the month that contains the seventh day after the ceremony. Unlike Study~1, the comparison rests on the shared nomination: the nominated albums come from the same eligibility year.

\subsection{Measures of the crowd and of the recordings played}

Four measures serve the tests of H3 and H4 in Study~1. The \emph{mainstream gap} compares the newcomers with the early fans on the measure of taste defined above. For the newcomers of months $0$ to $2$, we take the share of their plays in the six months before they discovered the artist that went to the year's 1,000 most-listened artists, and subtract the early fans' mean mainstream share. A positive gap marks a crowd whose listening is more mainstream than the early fans'. The \emph{size of the surge} is the newcomer ratio of the event: newcomers per month in months $0$ to $2$ relative to months $-12$ to $-1$. It counts only people playing the artist for the first time, so no early fan enters it. As in the analysis plan, we split the treated events into thirds by each measure and compare the top and bottom thirds.

For the recordings, we use the MusicBrainz recording that MLHD+ links to every play. The \emph{recording that drew the crowd} is the one that the largest number of newcomers of months $0$ to $2$ played in the month they discovered the artist. For each comparison artist we take, in the same way, the recording played by most of its own newcomers in those months. The artist's \emph{less-played recordings}, its deep cuts, are those outside its ten most-played recordings among all listeners in months $-12$ to $-1$. For each early fan who played the artist both in months $-12..{-7}$ and in months $0..5$, we compute the share of their plays of the artist that went to each kind of recording in both windows and take the change.

\subsection{Estimation}

Both studies compare each treated artist-event with the comparison artists of the same event, and we report the average over treated artist-events, so that every artist counts equally and a few very popular artists do not dominate. For the primary endpoint, we compute for each artist-event the share of its eligible early fans who stopped playing the artist (or came back), using artist-events with at least ten eligible fans; the effect is the treated artist's rate minus the mean rate of its comparison artists. The same contrast one window earlier is the placebo, and the difference between the two contrasts, computed within each event, tests whether the gap opens at the event. Because each transition starts from the fans who did or did not play the artist in the preceding window, the earlier and later contrasts compare transition rates of partly different sets of fans (Appendix~\ref{app:samples}). For the secondary outcomes, we compute each early fan's change in an outcome between a baseline and the months after the event (months $0$ to $12$, thirteen monthly bins), average these changes over the early fans of each artist and take the same treated-minus-comparison difference. We report three baselines: months $-12$ to $-7$ (the main specification), months $-6$ to $-1$ (which lie outside the selection window), and a linear trend fitted to the treated--comparison difference in months $-6$ to $-1$ and extrapolated through month $12$. The difference between the first two baselines is itself a placebo test: a treated--comparison difference that appears before the event points to diverging trends.

We report 95\% intervals from a bootstrap that resamples treated events together with their comparison artists (Study~1) or ceremonies (Study~2), and the range of estimates when each event is left out in turn. For the primary endpoint we also resample calendar months, so that artists exposed by the same film move together, and groups of events that share a comparison artist (Appendix~\ref{app:robust}). Artist-events with fewer than 30 early fans are dropped, and for the few very popular artists we analyse a random sample of 20,000 early fans. For the tests of H3, we average each treated event's quitting and returning gaps, and their change from the earlier window, over the events in each third of a measure, with intervals from resampling events within each third. For the recordings, each early fan's change in a share is averaged over the early fans of an artist-event (artist-events with at least ten such fans), and the treated minus comparison difference is averaged over events. For Study~2 we add a permutation benchmark that, in each ceremony and category, gives the award to a randomly chosen nominee. We also estimate fan-level regressions with fan fixed effects, event-by-window fixed effects and standard errors clustered by artist (Appendix~\ref{app:robust}).

\section{Results}
\label{sec:results}

\subsection{The crowd arrives}

Table~\ref{tab:samples} describes both studies. The \CrEvents crowd arrivals in Study~1 concern artists with an established audience: their early fans number \CrFansAllT in total, a median of \CrFansMedianT per artist, and \CrFansAllC early fans of \CrControlUnits comparison artist-events are the benchmark; after sampling, the analysis follows \CrUsersDistinct distinct users (Appendix~\ref{app:samples}). Before the event, the treated artist accounted for \CrShareRef\% of its early fans' plays, and early fans played it in \CrAnyRef\% of the months in which they used Last.fm. Treated and comparison artists were similar on most pre-event characteristics, including their early fans' tenure, taste and listening, the pre-event slope of the share of plays, and how often early fans stopped or came back in the months before the event (Appendix~\ref{app:balance}). Two differences are visible: treated artists had smaller audiences, and their newcomers had already begun to rise in the months before the surge.

\begin{figure}[t]
\centering
\includegraphics[width=\textwidth]{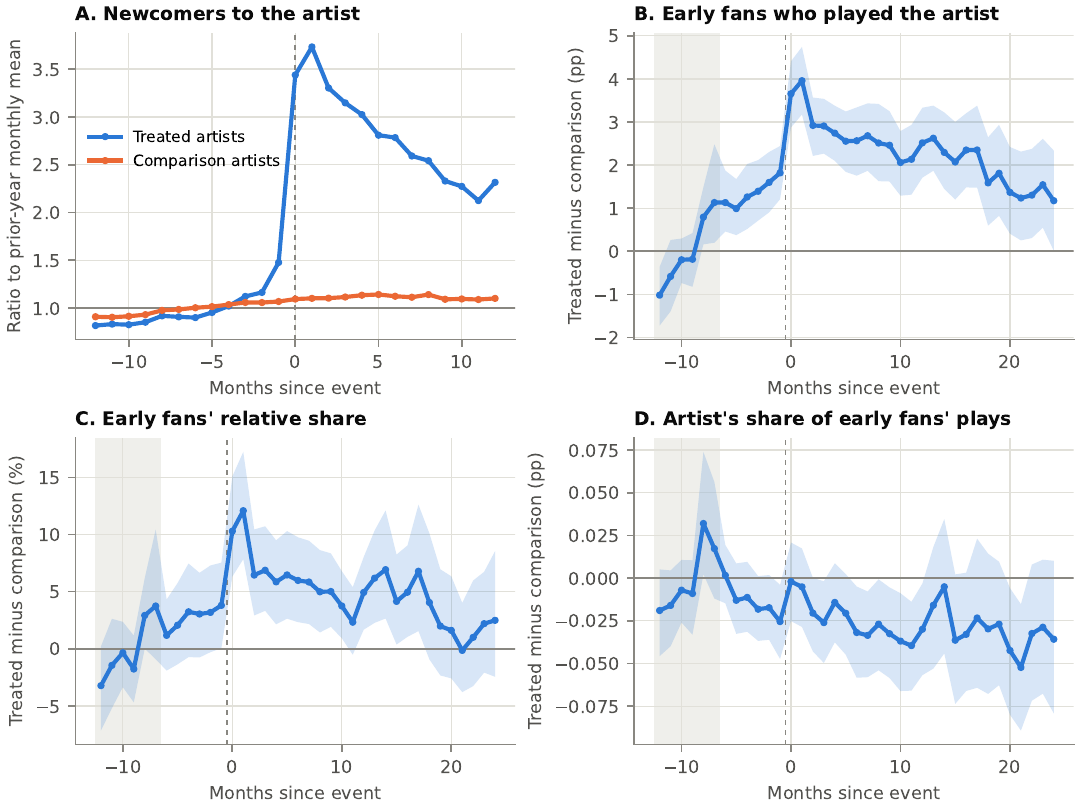}
\caption{Study 1, crowd arrivals, by month. Panel A: newcomers to treated and comparison artists, as the median ratio to each artist's monthly mean over the previous year. Panels B--D: treated minus comparison change in early fans' outcomes, relative to each fan's own average in months $-12$ to $-7$ (shaded), with 95\% bootstrap intervals over events. Months $-12$ to $-7$ fall inside the window used to select early fans; months $-6$ to $-1$ do not. All \CrEvents events contribute through month 12 and \CrEventsAtTwentyFour at month 24.}
\label{fig:crowd}
\end{figure}

The arrivals are large: over the event month and the two months after it, treated artists drew on average \CrNewRatioT times as many newcomers per month as in the previous year, against \CrNewRatioC times for comparison artists (Figure~\ref{fig:crowd}A). The surge decays but does not vanish: a year later newcomers still arrive at about twice the earlier rate. The number of people listening to the artist in a month rose by a median of \CrLisGrowthT\% over the first six months (interquartile range \CrLisGrowthQoneT\% to \CrLisGrowthQthreeT\%), against \CrLisGrowthC\% for comparison artists. For a typical treated artist, early fans found themselves sharing the artist with about as many new listeners as there had been listeners before. For \CrAgeOlderN of the \CrEvents artists the latest studio album was more than two years old at the time of the surge: the crowd arrived for an established catalogue, not for a first breakthrough. The largest arrivals (Appendix~\ref{app:events}) include Kavinsky and Desire, whose songs were used in the film \emph{Drive} (September 2011), and Blue Foundation and Anya Marina, whose songs appeared on the soundtracks of \emph{Twilight} (2008) and its sequel \emph{New Moon} (2009). They also include the Trashmen, whose 1963 single ``Surfin' Bird'' was featured in an October 2008 episode of the television series \emph{Family Guy}.

\begin{table}[t]
\centering
\small
\caption{The two studies. Early fans played the artist at least ten times in months $-24$ to $-7$ and used Last.fm in at least nine of the twelve months before the event. Newcomers are users who played the artist for the first time after at least six months of listening history; ratios add one to numerator and denominator.}
\label{tab:samples}
\begin{tabular}{p{0.55\textwidth}rr}
\toprule
 & Study 1: crowd arrivals & Study 2: award wins \\
\midrule
Events (stacks) & 78 & 9 \\
Treated artists & 78 & 17 \\
Comparison artists & 301 & 72 \\
Early fans of treated artists & 102,617 & 599,040 \\
Early fans of comparison artists & 606,175 & 1,862,573 \\
Early fans per treated artist (median) & 846 & 21{,}548 \\
Early fans analysed (after sampling) & 692,427 & 1,121,211 \\
Treated artist's share of early fans' plays, months $-12$ to $-7$ (\%) & 0.44 & 1.14 \\
Months in which early fans played the treated artist, months $-12$ to $-7$ (\%) & 39 & 49 \\
Newcomers in months 0--2, ratio to prior-year mean (median): treated artists & 3.53 & 1.20 \\
\quad comparison artists & 1.11 & 1.05 \\
Monthly listeners, months 0--5 vs $-12$ to $-7$ (median change, \%): treated artists & 123 & 81 \\
\quad comparison artists & 19 & 75 \\
\bottomrule
\end{tabular}
 
\end{table}

\subsection{Early fans quit less and return more (H1)}

\begin{figure}[t]
\centering
\includegraphics[width=\textwidth]{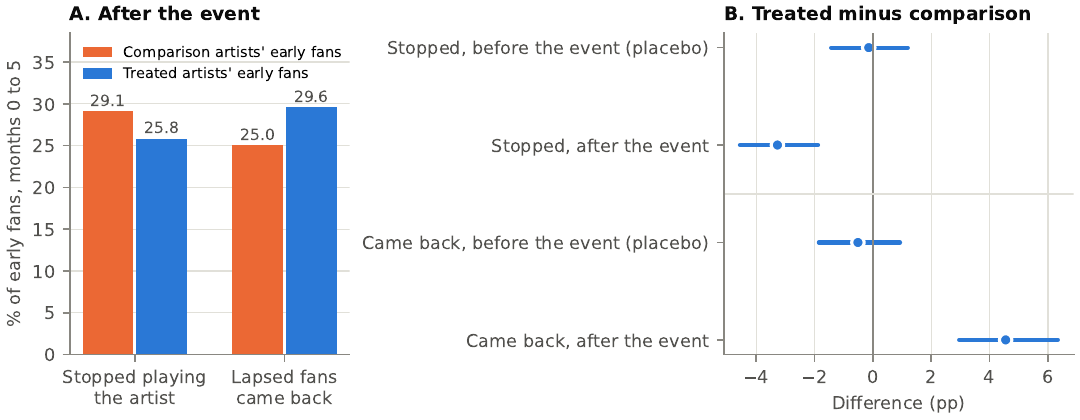}
\caption{Quitting and returning, Study 1. Panel A: share of early fans who played the artist in months $-6$ to $-1$ but not in months $0$ to $5$ (stopped), and share of those who had not played it in months $-6$ to $-1$ but did in months $0$ to $5$ (came back), for treated and comparison artists. Panel B: treated minus comparison differences with 95\% bootstrap intervals over events; the placebo rows repeat the comparison one window earlier (months $-12..{-7}$ to $-6..{-1}$). Fans must use Last.fm in at least four months of each window.}
\label{fig:transitions}
\end{figure}

The primary test of H1 is whether early fans stop playing the artist. They became less likely to do so, and lapsed fans more likely to return, as the renewed-attention account predicts (H1b) and against the identity account (H1a; Figure~\ref{fig:transitions}; Table~\ref{tab:main}, Panel~A). Among early fans who had played the artist in the six months before the event, \CrQuitT\% did not play it at all in the six months after, against \CrQuitC\% of comparison fans, a difference of \CrQuitDiff percentage points (95\% CI \CrQuitDifflo to \CrQuitDiffhi), or a rate \CrQuitRelPct\% lower. Among early fans who had drifted away, with no plays in the six months before the event, \CrReturnT\% came back in the six months after, against \CrReturnC\%, a difference of \CrReturnDiff points (95\% CI \CrReturnDifflo to \CrReturnDiffhi). Before the event there was no such gap: the same comparisons one window earlier give \CrQuitPlaceboDiff points for quitting (95\% CI \CrQuitPlaceboDifflo to \CrQuitPlaceboDiffhi) and \CrReturnPlaceboDiff for returning (95\% CI \CrReturnPlaceboDifflo to \CrReturnPlaceboDiffhi). Within each event, the quitting gap moved by \CrQuitChange points from the earlier window to the window after the event (95\% CI \CrQuitChangelo to \CrQuitChangehi), and the returning gap by \CrReturnChange points (95\% CI \CrReturnChangelo to \CrReturnChangehi). At the six-month resolution of these windows, the change coincides with the crowd arrival.

The result does not rest on any single event or choice. Leaving out one event at a time, the quitting difference stays between \CrQuitJackMin and \CrQuitJackMax points. Resampling calendar months instead of events, so that artists exposed by the same film move together, gives a 95\% interval of \CrQuitDiffMonthlo to \CrQuitDiffMonthhi points. The result holds for the stricter group of early fans who had known the artist for at least two years and played it in at least three separate months before (\CrStrictQuitDiff points, 95\% CI \CrStrictQuitDifflo to \CrStrictQuitDiffhi). Across \RobVariants alternative specifications the estimate ranges from \RobQuitMin to \RobQuitMax points, and none of the 95\% intervals includes zero (Appendix~\ref{app:robust}). With comparison artists chosen only on information up to month $-1$, it is \PastSameQuit points for the same \PastSameEvents events (95\% CI \PastSameQuitlo to \PastSameQuithi). Without every treated and comparison artist that had any new recording near the event, it is \AuditQuit points for the \AuditEvents events that remain (95\% CI \AuditQuitlo to \AuditQuithi), and \PastAuditQuit points (95\% CI \PastAuditQuitlo to \PastAuditQuithi) when both changes are made together. After the crowd arrived, fewer early fans went silent on the artist for six months than comparison fans did, and more of the lapsed ones came back.

\subsection{How much early fans listened}

The secondary outcomes describe how much early fans listened, and here the timing matters (Figure~\ref{fig:crowd}B--D; Table~\ref{tab:main}). The share of months in which early fans played the artist began to rise relative to the comparison before the event, as the artist's audience started to grow, jumped at the event and then declined slowly. Measured from months $-12$ to $-7$, it was \CrAnyPost percentage points higher over months $0$ to $12$ (95\% CI \CrAnyPostlo to \CrAnyPosthi); measured from the six months just before the event, \CrAnyPostPre points (95\% CI \CrAnyPostPrelo to \CrAnyPostPrehi). If the rise of the pre-event months is extrapolated as a straight line through month 12, the post-event difference is \CrAnyTrend points (95\% CI \CrAnyTrendlo to \CrAnyTrendhi). The relative share, which weighs light and heavy listeners equally, follows the same pattern: \CrRelPost\% (95\% CI \CrRelPostlo\% to \CrRelPosthi\%) from the earlier baseline, \CrRelPostPre\% (95\% CI \CrRelPostPrelo\% to \CrRelPostPrehi\%) from the six months before, and \CrRelTrend\% (95\% CI \CrRelTrendlo\% to \CrRelTrendhi\%) with the trend extrapolated.

The artist's share of early fans' total plays drifted the other way. It had been edging down relative to the comparison before the event (\CrSharePre percentage points in months $-6$ to $-1$, 95\% CI \CrSharePrelo to \CrSharePrehi) and kept doing so. From the earlier baseline it changed by \CrSharePost points over months $0$ to $12$ (\CrSharePct\% of the pre-event share; 95\% CI \CrSharePctlo\% to \CrSharePcthi\%). From the six months before it changed by \CrSharePostPre points (95\% CI \CrSharePostPrelo to \CrSharePostPrehi), and with the trend extrapolated by \CrShareTrend points (95\% CI \CrShareTrendlo to \CrShareTrendhi). The drift is largest among fans who had discovered the artist most recently, consistent with newer fans' interest fading on its own (Appendix~\ref{app:robust}).

Taken together, the listening measures show no break towards abandonment at the event. Early fans played the artist in more months, partly continuing a rise that had begun before, and the artist took a slowly declining share of their total plays, continuing an earlier drift. In the second year, months 13 to 24, early fans still played the artist in more months than comparison fans (\CrAnyLong points, 95\% CI \CrAnyLonglo to \CrAnyLonghi; \CrEventsAtTwentyFour events contribute at month 24), while the artist's share of their plays remained slightly lower (\CrShareLong points, 95\% CI \CrShareLonglo to \CrShareLonghi).

\begin{table}[t]
\centering
\small
\caption{Main effects. Panel A: treated minus comparison difference in the share of early fans who stopped playing the artist (among those who had played it) or came back (among those who had not), from one six-month window to the next. Panel B: change in early fans' listening from the baseline shown to months $0$ to $12$, treated minus comparison, averaged over artist-events with each weighted equally; the trend row extrapolates the treated--comparison difference of months $-6$ to $-1$ linearly. 95\% bootstrap intervals over events (Study 1) and ceremonies (Study 2) in brackets.}
\label{tab:main}
\setlength{\tabcolsep}{4pt}
\begin{tabular}{llcc}
\toprule
Outcome & Comparison & Study 1: crowd arrivals & Study 2: award wins \\
\midrule
\multicolumn{4}{l}{\emph{A. Primary endpoint: stopping and returning (percentage points)}} \\
Stopped playing the artist & Months $-6..{-1}$ to $0..5$ & $-$3.3 [$-$4.6, $-$1.9] & $-$2.1 [$-$4.9, 0.3] \\
 & Months $-12..{-7}$ to $-6..{-1}$ (placebo) & $-$0.1 [$-$1.4, 1.2] & $-$1.7 [$-$5.1, 1.1] \\
Lapsed fans came back & Months $-6..{-1}$ to $0..5$ & 4.6 [3.0, 6.3] & 2.5 [0.2, 5.1] \\
 & Months $-12..{-7}$ to $-6..{-1}$ (placebo) & $-$0.5 [$-$1.9, 0.9] & 4.3 [0.4, 8.3] \\
\addlinespace
\multicolumn{4}{l}{\emph{B. Listening (change in early fans, treated minus comparison)}} \\
Listened at all (pp) & Months $0..12$ vs $-12..{-7}$ & 2.8 [2.1, 3.4] & 1.9 [$-$2.3, 6.2] \\
 & Months $0..12$ vs $-6..{-1}$ & 1.4 [0.8, 1.9] & $-$0.4 [$-$2.6, 1.9] \\
 & Months $0..12$, pre-trend extrapolated & $-$0.1 [$-$1.2, 1.1] & $-$0.9 [$-$5.4, 3.4] \\
 & Months $-6..{-1}$ vs $-12..{-7}$ (placebo) & 1.3 [0.8, 1.9] & 2.3 [$-$2.5, 6.8] \\
 & Months $13..24$ vs $-12..{-7}$ & 1.8 [1.0, 2.6] & 0.2 [$-$4.1, 5.4] \\
\addlinespace
Relative share (\%) & Months $0..12$ vs $-12..{-7}$ & 6.2 [3.3, 9.1] & 5.5 [$-$10.6, 24.8] \\
 & Months $0..12$ vs $-6..{-1}$ & 3.5 [1.3, 5.9] & 0.2 [$-$11.5, 14.2] \\
 & Months $0..12$, pre-trend extrapolated & $-$1.0 [$-$7.1, 5.6] & 0.4 [$-$31.0, 34.0] \\
 & Months $-6..{-1}$ vs $-12..{-7}$ (placebo) & 2.7 [$-$0.4, 5.8] & 5.3 [$-$20.6, 30.3] \\
 & Months $13..24$ vs $-12..{-7}$ & 3.6 [0.5, 6.8] & $-$2.1 [$-$17.1, 12.3] \\
\addlinespace
Share of plays (pp) & Months $0..12$ vs $-12..{-7}$ & $-$0.024 [$-$0.047, $-$0.002] & $-$0.035 [$-$0.269, 0.186] \\
 & Months $0..12$ vs $-6..{-1}$ & $-$0.010 [$-$0.028, 0.005] & $-$0.068 [$-$0.209, 0.069] \\
 & Months $0..12$, pre-trend extrapolated & 0.031 [$-$0.002, 0.064] & 0.047 [$-$0.427, 0.482] \\
 & Months $-6..{-1}$ vs $-12..{-7}$ (placebo) & $-$0.014 [$-$0.030, 0.002] & 0.033 [$-$0.223, 0.311] \\
 & Months $13..24$ vs $-12..{-7}$ & $-$0.032 [$-$0.065, $-$0.001] & $-$0.112 [$-$0.340, 0.097] \\
\addlinespace
\multicolumn{2}{l}{Treated share of plays, months $-12..{-7}$ (\%)} & 0.44 & 1.14 \\
\multicolumn{2}{l}{Treated artist-events} & 78 & 18 \\
\bottomrule
\end{tabular}
 
\end{table}

\subsection{Who stays (H2)}

\begin{figure}[t]
\centering
\includegraphics[width=\textwidth]{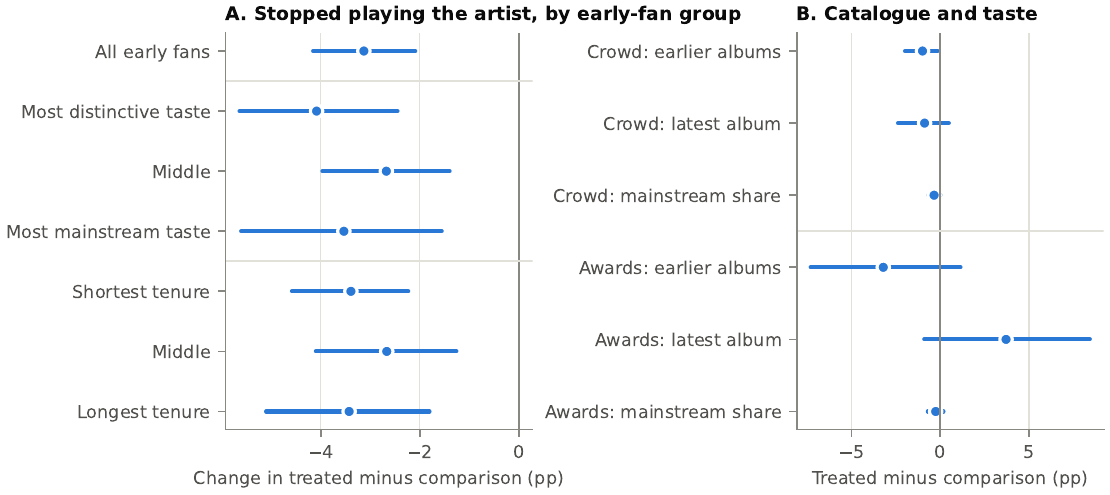}
\caption{Who stays and what early fans play. Panel A: change in the treated minus comparison difference in the share of early fans who stopped playing the artist, from the earlier window (months $-12..{-7}$ to $-6..{-1}$) to the window after the event (months $-6..{-1}$ to $0..5$), paired within event, for all early fans and for thirds by the mainstream share of their other listening and by tenure with the artist. Panel B: change in the share of early fans' plays of the artist going to its earlier studio albums and to its latest studio album (for artist-events whose latest album was the same throughout months $-12$ to $12$), and in the share of their other listening going to the year's 1,000 most popular artists. Lines are 95\% bootstrap intervals over events.}
\label{fig:mech}
\end{figure}

If the crowd's arrival imposed an identity cost, it should weigh most on fans whose tastes lie furthest from the mainstream (H2a). If it renewed early fans' attention to the artist, the change should be largest among fans whose listening had lapsed or thinned, whatever their taste (H2b). We split early fans into thirds by their mainstream share, the share of their other listening that went to the year's 1,000 most popular artists. After the event, the treated minus comparison difference in quitting was negative in all three thirds. It was \CrMsQuitA percentage points (95\% CI \CrMsQuitAlo to \CrMsQuitAhi) among the third with the most distinctive tastes (a mainstream share of at most \CrMsCutLo\%) and \CrMsQuitC points (95\% CI \CrMsQuitClo to \CrMsQuitChi) among the most mainstream third (above \CrMsCutHi\%). The difference between these two groups, \CrMsQuitContrast points, was already there before the event: in the earlier window it was \CrMsQuitPlaceboContrast points (95\% CI \CrMsQuitPlaceboContrastlo to \CrMsQuitPlaceboContrasthi), so the gradient predates the crowd. From the earlier window to the window after the event, this difference changed by \CrMsQuitChangeA points among the most distinctive third (95\% CI \CrMsQuitChangeAlo to \CrMsQuitChangeAhi), \CrMsQuitChangeB among the middle third and \CrMsQuitChangeC among the most mainstream third (95\% CI \CrMsQuitChangeClo to \CrMsQuitChangeChi; Figure~\ref{fig:mech}A). The difference between the first and last of these changes is \CrMsQuitContrastChange points (95\% CI \CrMsQuitContrastChangelo to \CrMsQuitContrastChangehi), and returning shows the same pattern (\CrMsReturnContrastChange points, 95\% CI \CrMsReturnContrastChangelo to \CrMsReturnContrastChangehi). The crowd's arrival did not move fans with distinctive tastes differently from fans with mainstream tastes. The listening measures give no consistent gradient either. The relative share rose less among the most distinctive third than among the most mainstream third (a difference of \CrMsRelContrast points, 95\% CI \CrMsRelContrastlo to \CrMsRelContrasthi), while the artist's share of plays fell less among them (a difference of \CrMsShareContrast points, 95\% CI \CrMsShareContrastlo to \CrMsShareContrasthi).

Listening histories mattered more, in the way H2b predicts, for returning and for how much fans listened, though not for quitting. Returning rose most among the lapsed fans who had found the artist most recently. From the earlier window to the window after the event, the returning difference changed by \CrTenReturnChangeA points among the third with the shortest tenure, against \CrTenReturnChangeC for the longest (a difference of \CrTenReturnContrastChange points, 95\% CI \CrTenReturnContrastChangelo to \CrTenReturnContrastChangehi). The increase in listening came mostly from light listeners: their relative share rose by \CrIntRelA\% (95\% CI \CrIntRelAlo\% to \CrIntRelAhi\%), against \CrIntRelC\% (95\% CI \CrIntRelClo\% to \CrIntRelChi\%) for the heaviest. Quitting, by contrast, fell by similar amounts across tenure (a change of \CrTenQuitChangeA points among the shortest third and \CrTenQuitChangeC among the longest; difference \CrTenQuitContrastChange points, 95\% CI \CrTenQuitContrastChangelo to \CrTenQuitContrastChangehi) and across listening intensity (\CrIntQuitA points among the lightest listeners and \CrIntQuitC among the heaviest after the event). The longest-standing fans, for whom having been there first should be worth most, gained the least on returning, which does not fit reflected glory as the main reason for staying.

\subsection{The size and makeup of the crowd (H3)}

The size of the surge separated the accounts (Table~\ref{tab:mechanism}). We split the treated events into thirds by their newcomer ratio, which counts only people playing the artist for the first time (cut points \MechNrCutLo and \MechNrCutHi). From the earlier window to the window after the event, the returning gap rose by \MechNrRetChgLo points in the third with the smallest surges (95\% CI \MechNrRetChgLolo to \MechNrRetChgLohi), by \MechNrRetChgMid in the middle third and by \MechNrRetChgHi in the third with the largest (95\% CI \MechNrRetChgHilo to \MechNrRetChgHihi). The difference between the top and bottom thirds is \MechNrRetChgDiff points (95\% CI \MechNrRetChgDifflo to \MechNrRetChgDiffhi). Quitting did not rise with the size of the surge; if anything it fell more (a difference of \MechNrQuitChgDiff points, 95\% CI \MechNrQuitChgDifflo to \MechNrQuitChgDiffhi). More lapsed fans returned after larger surges, as renewed attention predicts (H3b), and larger surges were not followed by more quitting (H3a). Splitting by the growth in monthly listeners instead, a measure that also counts the returning fans, gives a returning difference of \MechGrowRetChgDiff points (95\% CI \MechGrowRetChgDifflo to \MechGrowRetChgDiffhi).

The crowds were, on the whole, close to the early fans in how mainstream their listening was. Across the \MechEvents events, the mainstream gap (the newcomers' mainstream share minus the early fans') had a median of \MechDisMedian points (interquartile range \MechDisQone to \MechDisQthree), and in \MechDisPosN events the crowd's listening was more mainstream than the early fans'. If identity costs drive early fans away, quitting should have risen most where the crowd was more mainstream than they were (H3a). It did not. From the earlier window to the window after the event, the quitting gap changed by \MechDisQuitChgHi points (95\% CI \MechDisQuitChgHilo to \MechDisQuitChgHihi) in the third of events whose crowds were furthest towards the mainstream relative to their early fans (a gap above \MechDisCutHi points). It changed by \MechDisQuitChgLo points (95\% CI \MechDisQuitChgLolo to \MechDisQuitChgLohi) in the third with the smallest or negative gaps, a difference of \MechDisQuitChgDiff points (95\% CI \MechDisQuitChgDifflo to \MechDisQuitChgDiffhi). Returning did not differ either (\MechDisRetChgDiff points, 95\% CI \MechDisRetChgDifflo to \MechDisRetChgDiffhi). After the event alone, the quitting gap was smaller where the crowd was more mainstream (\MechDisQuitHi against \MechDisQuitLo points), but, as with the fans' own tastes, that difference was already there before the crowd arrived.

\begin{table}[t]
\centering
\small
\caption{Tests of the crowd's size and makeup (H3) and of the recordings early fans played (H4), Study~1. Quitting and returning: difference between the top and bottom thirds of treated events, by each measure, in the change of the treated minus comparison gap from the earlier window (months $-12..{-7}$ to $-6..{-1}$) to the window after the event (months $-6..{-1}$ to $0..5$), in percentage points. The size of the surge is the newcomer ratio; the mainstream gap is the newcomers' mainstream share minus the early fans'. Recordings: treated minus comparison change in the share of early fans' plays of the artist from months $-12..{-7}$ to $0..5$, in percentage points. 95\% bootstrap intervals over events in brackets.}
\label{tab:mechanism}
\begin{tabularx}{\textwidth}{>{\raggedright\arraybackslash}X>{\raggedright\arraybackslash}p{0.13\textwidth}>{\raggedright\arraybackslash}p{0.15\textwidth}r}
\toprule
Test & Identity account & Renewed attention & Estimate [95\% CI] \\
\midrule
Quitting: crowd more mainstream than the early fans, top minus bottom third & Higher & No difference & $-$0.2 [$-$2.4, 2.2] \\
Returning: same split & Lower & No difference & 0.6 [$-$3.1, 4.6] \\
Quitting: largest minus smallest third of surges & Higher & Lower & $-$1.3 [$-$4.1, 1.4] \\
Returning: same split & Lower & Higher & 6.2 [1.6, 11.2] \\
Share of plays: the recording that drew the crowd & Falls & Rises & 3.0 [2.1, 4.0] \\
Share of plays: the artist's less-played recordings & Rises & No prediction & 0.3 [$-$0.6, 1.2] \\
\bottomrule
\end{tabularx}
 
\end{table}

\subsection{What early fans played (H4)}

Early fans did not shift towards the artist's earlier albums (H4a; Figure~\ref{fig:mech}B). MLHD+ links each play to a canonical release of its recording, usually the album on which it first appeared, and we classify plays by that release: the artist's latest studio album, its earlier studio albums, or anything else. We fix the categories at the event, keeping the \CrOlderNevents treated artist-events whose latest studio album stayed the same from month $-12$ to month $12$. Among early fans' plays of the artist, the share from earlier studio albums changed by \CrOlderPost percentage points relative to the comparison (95\% CI \CrOlderPostlo to \CrOlderPosthi) from \CrOlderRef\%, and the share from the latest studio album by \CrLatestPost points (95\% CI \CrLatestPostlo to \CrLatestPosthi). The remaining share, which covers singles, EPs, live albums and compilations as well as releases credited to other artists such as soundtracks (Appendix~\ref{app:data}), rose by \CrOtherPost points (95\% CI \CrOtherPostlo to \CrOtherPosthi). Early fans' listening to other artists barely moved: the share going to the year's most popular artists changed by \CrMainstreamPost points (95\% CI \CrMainstreamPostlo to \CrMainstreamPosthi) from \CrMainstreamRef\%, and the rate at which they tried artists new to them by \CrDiscoveryPost per 100 plays (95\% CI \CrDiscoveryPostlo to \CrDiscoveryPosthi).

The recordings point the same way (Table~\ref{tab:mechanism}). In \MechRecCueNotTopN of the \MechRecEvents treated events, the recording that most newcomers played in the month they discovered the artist was not the artist's most-played recording of the previous year: the crowd usually came for a particular song. Among early fans who played the artist both before and after the event, the share of their plays that went to this recording rose by \MechRecCuePost percentage points relative to the comparison (95\% CI \MechRecCuePostlo to \MechRecCuePosthi), from \MechRecCueRefT\%. It had not been rising before the event (\MechRecCuePre points from months $-12..{-7}$ to $-6..{-1}$, 95\% CI \MechRecCuePrelo to \MechRecCuePrehi), and it was still higher in months 6 to 12 (\MechRecCueLate points, 95\% CI \MechRecCueLatelo to \MechRecCueLatehi). The rise is not an artefact of comparing a newly popular song with comparison artists' established hits. The share rose by \MechRecCueRawT points among treated early fans and by \MechRecCueRawC among comparison fans. In the \MechRecCueTopOnlyN events whose crowd came for the artist's most-played recording, set against comparison artists whose newcomers also chose theirs, it rose by \MechRecCueTopOnly points (95\% CI \MechRecCueTopOnlylo to \MechRecCueTopOnlyhi). The share going to the artist's less-played recordings, those outside its ten most-played, changed by \MechRecDeepPost points (95\% CI \MechRecDeepPostlo to \MechRecDeepPosthi) from \MechRecDeepRefT\%. For most early fans who played the song in the six months after the event, it was not new: \MechRecCueKnownPctT\% of them had already played it in the year before (\MechRecCueKnownPctC\% for comparison artists' fans). The rest played it for the first time in the record. Early fans who kept listening converged on the song that drew the crowd (H4b) instead of turning away from it to the deep cuts (H4a).

\subsection{The crowd itself}

The newcomers who came with a surge kept listening about as often as those who had found the artist earlier. We compare listeners who discovered an artist in the three months of the surge with those who discovered it in months $-24$ to $-13$, whose own three-to-twelve-month follow-up ends before the event. Both groups are limited to newcomers who kept using Last.fm (\CrRetNsurgeT surge and \CrRetNrefT earlier newcomers to treated artists; artist-events with fewer than \CrRetMinRef earlier newcomers are left out). Of the surge arrivals who kept using Last.fm, \CrRetSurgeT\% played the artist again three to twelve months later, against \CrRetRefT\% of the earlier arrivals; for comparison artists the figures were \CrRetSurgeC\% and \CrRetRefC\%. Relative to that benchmark, retention among surge arrivals differed by \CrRetDiD percentage points (95\% CI \CrRetDiDlo to \CrRetDiDhi): arriving with the crowd carried no detectable retention penalty. The surge newcomers' tastes were, on average, about as mainstream as the early fans': \CrNewMsSurgeT\% of their listening in the six months before they discovered the artist went to the year's most popular artists, against \CrFanMsT\% of the early fans' listening in the selection window.

The crowd changed the artist's audience as a whole. Counting all of its listeners, the artist's plays per listener fell by \MechPplPctDiffAbs\% relative to comparison artists from months $-12..{-7}$ to months $0..5$ (95\% CI \MechPplPctDiffhiAbs\% to \MechPplPctDiffloAbs\%). An average taken over the whole audience thus declined while the early fans quit less often than comparison fans. The two statistics measure different behaviours over different denominators, and a falling average over the audience is no evidence that the first listeners left.

\subsection{Study 2: award wins}

Awards produce a sharp and public status shock, but on Last.fm they produced little additional crowd. Winners and losing nominees both gained listeners in the months around a ceremony, mostly from the albums that earned the nominations. In the ceremony month and the two after, winners drew \AwNewRatioT times their prior-year newcomers against \AwNewRatioC times for losing nominees, and their monthly listeners grew by \AwLisGrowthT\% against \AwLisGrowthC\% (Table~\ref{tab:samples}; Figure~\ref{fig:awards} in the Appendix). The study therefore asks a narrower question than Study~1: does winning, as such, change how existing fans listen?

The comparison covers \AwNevents winner artist-events from \AwCeremonies ceremonies (\AwWinnerArtists distinct artists). The quitting difference between winners' and losing nominees' early fans was \AwQuitDiff percentage points (95\% CI \AwQuitDifflo to \AwQuitDiffhi), against \AwQuitPlaceboDiff points one window earlier; returning differed by \AwReturnDiff points, but a gap of similar size had existed before the ceremony (\AwReturnPlaceboDiff points). The artist's share of early fans' plays changed by \AwSharePost percentage points (95\% CI \AwSharePostlo to \AwSharePosthi) from \AwShareRef\%, an interval that includes declines of up to \AwSharePctloAbs\% as well as increases. Giving the award to a randomly chosen nominee in each category produces a share difference at least this large in \AwPermSharePct\% of draws. Leaving out one ceremony at a time, the share estimate ranges from \AwLooShareMin to \AwLooShareMax points. The award study thus finds no clear change in winners' early fans.

\section{Discussion}
\label{sec:discussion}

\subsection{Summary}

Identity-signalling theory, and the stock phrase that echoes it, expect an artist's first fans to pull away once the artist becomes popular. In the listening histories of several hundred thousand Last.fm users, the behaviour this predicts did not appear, and its opposite did. When an artist's audience roughly doubled within months without a new release, fewer of its early fans stopped playing it and more of those who had drifted away came back (H1b, not H1a). Early fans who kept listening converged on the crowd: a larger share of their plays went to the song most newcomers played, and their plays did not move towards the deep cuts or the earlier albums (H4b, not H4a). The change did not differ between fans with niche and with mainstream tastes; it was clearest, for returning and for how much fans listened, among fans whose listening had lapsed or thinned (H2b, not H2a). Quitting did not rise where the newcomers were more mainstream than the early fans, and the larger the surge, the more lapsed fans returned (H3b, not H3a). The newcomers who came with the crowd kept listening about as often as earlier newcomers. Award wins, which brought winners few additional listeners on Last.fm, produced no clear change in their early fans.

\subsection{Renewed attention, not withdrawal}

The pattern is the one renewed attention predicts. The response sits at the margin of whether early fans play the artist at all, not at how much the devoted play it, and it is clearest among fans for whom the artist had left their regular rotation. It grows with the number of people the event reached, it jumped at the event and then declined slowly, and it concentrates on the particular song the crowd came for, which most early fans who played it already knew. Each of these follows from the event reaching the people who already knew the artist as well as the newcomers, as concerts, chart entries and film placements do \citep{Ternovski2020,Sim2022,Aguiar2024}.

The marks of identity-driven distancing are missing. Identity-signalling theory expects withdrawal to weigh most on fans whose tastes are the most distinctive and to grow when the adopters are a group the early fans do not want to be mistaken for \citep{Berger2008}. Neither gradient appeared. Nor did the fans who stayed distinguish themselves within the artist's work: they played more of the crowd's song, not less, and no more of the deep cuts. The longest-standing fans, for whom the status of having been there first should be worth most, gained the least on returning, so reflected glory \citep{Cialdini1976} does not explain the pattern either.

The results concern behaviour, not feelings. A fan can replay a song she heard in a film, resent the newcomers and tell friends she found the band years ago. Identity motives can also favour staying, since the claim to have been there first requires having been a fan, and such claims live in conversation and in deliberate displays. What the results show is that whatever identity costs early fans felt did not translate into withdrawal from the artist or from the crowd's song in a record that their profiles displayed by default. That is a strong setting for the identity account to show up in behaviour. Last.fm users are known to change what they play to manage their profiles \citep{Silfverberg2011}, and identity motives have been shown to shape what people choose to share \citep{Johnson2018}. Posting about a product can satisfy the need to signal an identity and lower later interest in buying it \citep{Grewal2019}, which suggests that the identity work may be done by what people deliberately show rather than by what they play.

\subsection{Implications for theory}

The results change three claims. First, identity-signalling theory predicts that people abandon a taste once others adopt it \citep{Berger2008}, and research on online taste performance has shown identity motives at work in what people choose to share \citep{Johnson2018}. The theory's authors already expected people to keep tastes they like in private while diverging in public \citep[p.~598]{Berger2008}. Our evidence concerns the case in between: consumption chosen for oneself that leaves a public, persistent trace by default. There, a sudden spread of the taste was followed by less quitting, more returning and convergence on the song that drew the crowd, with no change by taste distinctiveness or by how mainstream the crowd was. A large influx of adopters is not sufficient to produce behavioural withdrawal from a held taste, even when the behaviour is displayed. The abandonment prediction should be scoped to deliberate displays, such as sharing, featuring or recommending, and to choices made in view of a known, dissimilar group.

Second, research on how public events raise demand for music has estimated their effect on streams, discovery and repeat consumption \citep{Ternovski2020,Sim2022,Aguiar2024}, and research in this journal has shown that audience size shapes how people experience content \citep{Waddell2020}. Following identifiable prior listeners adds who in the existing audience responds and how. The response came mostly from fans who had drifted away and from light and recent listeners, it grew with the size of the surge, and it went to the recording the crowd came for. For the people who were there first, a sudden audience worked as a renewal of attention, not as a reason to leave.

Third, research on status shocks explains the fall in average evaluations after an award as a mix of a new audience and a reaction against popularity \citep{Kovacs2014}. Following individuals across an influx shows how far the audience's average and the first audience's behaviour can come apart. The artist's plays per listener fell after the crowd arrived, the kind of drop that reads as a cooling audience, while its existing listeners quit less. Without comparison artists, the artist's share of its early fans' listening fell by \CrRawSharePctAbsT\% in the year after a crowd arrival, which would read as an exodus. Yet the early fans of comparison artists, selected in the same way and at the same point of the release cycle, reduced it by \CrRawSharePctAbsC\%. Listening to any artist rises and decays around each release, and fans selected for heavy listening in one period listen less in the next. Following the same people over time, and comparing artists at the same point of their release cycle, separates composition from behaviour and these artefacts from abandonment.

\subsection{Implications for practice}

For artists, labels and platforms, a surge of attention brought on by a film, a television series or a viral moment did not, on average, cost an artist its existing audience in what that audience plays. It is a moment to reach lapsed listeners: the fans who returned were mostly those who had drifted away, more of them returned when the surge was larger, and they played the song that drew the crowd. A platform could notify listeners when an artist they once played becomes popular, or surface that song to those who already know the artist. Today's services turn listening into public statements about people, as Spotify's yearly summaries do \citep{Annabell2026}; our results suggest that such displays need not drive early listeners away from an artist whose audience grows. Falling plays per listener after a breakthrough should not be read as a cooling base; the measure mixes the newcomers with the people who were there first, and only following the same listeners shows what the first audience did.

\subsection{Limitations and future work}

The data come from Last.fm users between 2005 and 2013, a population of committed music enthusiasts who chose to log their listening, and include only the plays that could be matched to MusicBrainz recordings. Future work can test whether listeners on today's streaming services, where algorithmic playlists shape much of what people hear, respond in the same way. The comparison artists are matched on the release cycle and audience size, and the result holds when they are chosen with information from before the event only, but crowd arrivals are not randomly assigned. Designs that use plausibly random exposure, such as the timing of a broadcast across regions, could estimate the causal effect more sharply. The same designs could separate the routes that renewed attention joins: hearing the song again, seeing that others now listen, and a radio station or playlist supplying the play. The crowd arrivals we study are surges without new albums; how early fans respond when a breakthrough comes with a new and different-sounding album, where a change in the music and a change in the audience coincide, remains open. The award comparison rests on nominees of the same year, who can be at different stages of their careers, such as a Best New Artist nominee next to an established band. More ceremonies and prizes would allow award winners to be matched on the release cycle as in Study~1.

Several measures can be refined. Many artists in both groups had one or two new recordings near the event, and the album categories rest on the canonical release of each recording. Our measure of the crowd counts newcomers on Last.fm; linking it to sales, radio play or social media attention would describe the wider audience that early fans see. The version of the data we use holds no information on users' countries, so early fans who saw a ceremony or a television show cannot be compared with those who did not. Our measures of distinctive taste and of the crowd's makeup are behavioural and rest on one dimension, how much of a person's listening goes to the most popular artists, and most crowds were close to the early fans on it. Crowds that differ in genre or in social identity, and survey measures of the need for uniqueness and of identification with an artist, would test the identity account where it should be strongest. Finally, we observe listening, not the tastes people choose to display. An experiment could give people who already listen to an artist the same reminder of the same song and vary whether the new adopters are a group they would not want to be mistaken for. It could then offer two choices: replaying the song, which enters their visible history by default, and featuring it on their profile. The identity account in the scope proposed here predicts a larger penalty for featuring than for replaying.

\subsection{Conclusion}

Early fans do not abandon artists who suddenly become popular, at least not in what they play. When the crowd arrives, the people who were there first move with it: fewer stop listening, more of those who had drifted away return, more of them the larger the surge, and those who keep listening play more of the song the newcomers came for. The hipster's lament may be real as talk, but in a listening record that every profile displays, the first listeners moved towards the artist after the crowd arrived, not away from it.

\bibliographystyle{plainnat}
\bibliography{references}

\begin{thebibliography}{37}
\providecommand{\natexlab}[1]{#1}
\providecommand{\url}[1]{\texttt{#1}}
\expandafter\ifx\csname urlstyle\endcsname\relax
  \providecommand{\doi}[1]{doi: #1}\else
  \providecommand{\doi}{doi: \begingroup \urlstyle{rm}\Url}\fi

\bibitem[Aguiar and Chen(2024)]{Aguiar2024}
Luis Aguiar and Zhizhong Chen.
\newblock Let that sync in: The effect of music reuse on product discovery.
\newblock SSRN working paper 4922422, 2024.

\bibitem[Anderson et~al.(2014)Anderson, Kumar, Tomkins, and
  Vassilvitskii]{Anderson2014}
Ashton Anderson, Ravi Kumar, Andrew Tomkins, and Sergei Vassilvitskii.
\newblock The dynamics of repeat consumption.
\newblock In \emph{Proceedings of the 23rd International Conference on World
  Wide Web}, pages 419--430. ACM, 2014.
\newblock \doi{10.1145/2566486.2568018}.

\bibitem[Annabell and Rasmussen(2026)]{Annabell2026}
Taylor Annabell and Nina~Vindum Rasmussen.
\newblock An algorithmic event: The celebration and critique of {Spotify
  Wrapped}.
\newblock \emph{New Media \& Society}, 28\penalty0 (10):\penalty0 5620--5643,
  2026.
\newblock \doi{10.1177/14614448251391301}.

\bibitem[Arsel and Thompson(2011)]{Arsel2011}
Zeynep Arsel and Craig~J. Thompson.
\newblock Demythologizing consumption practices: How consumers protect their
  field-dependent identity investments from devaluing marketplace myths.
\newblock \emph{Journal of Consumer Research}, 37\penalty0 (5):\penalty0
  791--806, 2011.
\newblock \doi{10.1086/656389}.

\bibitem[Baym and Ledbetter(2009)]{Baym2009}
Nancy~K. Baym and Andrew Ledbetter.
\newblock Tunes that bind? {P}redicting friendship strength in a music-based
  social network.
\newblock \emph{Information, Communication \& Society}, 12\penalty0
  (3):\penalty0 408--427, 2009.
\newblock \doi{10.1080/13691180802635430}.

\bibitem[Berger and Heath(2007)]{Berger2007}
Jonah Berger and Chip Heath.
\newblock Where consumers diverge from others: Identity signaling and product
  domains.
\newblock \emph{Journal of Consumer Research}, 34\penalty0 (2):\penalty0
  121--134, 2007.
\newblock \doi{10.1086/519142}.

\bibitem[Berger and Heath(2008)]{Berger2008}
Jonah Berger and Chip Heath.
\newblock Who drives divergence? {I}dentity signaling, outgroup dissimilarity,
  and the abandonment of cultural tastes.
\newblock \emph{Journal of Personality and Social Psychology}, 95\penalty0
  (3):\penalty0 593--607, 2008.
\newblock \doi{10.1037/0022-3514.95.3.593}.

\bibitem[Berger and Le~Mens(2009)]{Berger2009}
Jonah Berger and Ga{\"e}l Le~Mens.
\newblock How adoption speed affects the abandonment of cultural tastes.
\newblock \emph{Proceedings of the National Academy of Sciences}, 106\penalty0
  (20):\penalty0 8146--8150, 2009.
\newblock \doi{10.1073/pnas.0812647106}.

\bibitem[Berger and Ward(2010)]{Berger2010}
Jonah Berger and Morgan Ward.
\newblock Subtle signals of inconspicuous consumption.
\newblock \emph{Journal of Consumer Research}, 37\penalty0 (4):\penalty0
  555--569, 2010.
\newblock \doi{10.1086/655445}.

\bibitem[Bourdieu(1984)]{Bourdieu1984}
Pierre Bourdieu.
\newblock \emph{Distinction: A Social Critique of the Judgement of Taste}.
\newblock Harvard University Press, Cambridge, MA, 1984.

\bibitem[Brewer(1991)]{Brewer1991}
Marilynn~B. Brewer.
\newblock The social self: On being the same and different at the same time.
\newblock \emph{Personality and Social Psychology Bulletin}, 17\penalty0
  (5):\penalty0 475--482, 1991.
\newblock \doi{10.1177/0146167291175001}.

\bibitem[Bryson(1996)]{Bryson1996}
Bethany Bryson.
\newblock ``{A}nything but heavy metal'': Symbolic exclusion and musical
  dislikes.
\newblock \emph{American Sociological Review}, 61\penalty0 (5):\penalty0
  884--899, 1996.
\newblock \doi{10.2307/2096459}.

\bibitem[Chan et~al.(2012)Chan, Berger, and Van~Boven]{Chan2012}
Cindy Chan, Jonah Berger, and Leaf Van~Boven.
\newblock Identifiable but not identical: Combining social identity and
  uniqueness motives in choice.
\newblock \emph{Journal of Consumer Research}, 39\penalty0 (3):\penalty0
  561--573, 2012.
\newblock \doi{10.1086/664804}.

\bibitem[Cialdini et~al.(1976)Cialdini, Borden, Thorne, Walker, Freeman, and
  Sloan]{Cialdini1976}
Robert~B. Cialdini, Richard~J. Borden, Avril Thorne, Marcus~Randall Walker,
  Stephen Freeman, and Lloyd~Reynolds Sloan.
\newblock Basking in reflected glory: Three (football) field studies.
\newblock \emph{Journal of Personality and Social Psychology}, 34\penalty0
  (3):\penalty0 366--375, 1976.
\newblock \doi{10.1037/0022-3514.34.3.366}.

\bibitem[Grewal et~al.(2019)Grewal, Stephen, and Coleman]{Grewal2019}
Lauren Grewal, Andrew~T. Stephen, and Nicole~Verrochi Coleman.
\newblock When posting about products on social media backfires: The negative
  effects of consumer identity signaling on product interest.
\newblock \emph{Journal of Marketing Research}, 56\penalty0 (2):\penalty0
  197--210, 2019.
\newblock \doi{10.1177/0022243718821960}.

\bibitem[Johnson and Ranzini(2018)]{Johnson2018}
Benjamin~K. Johnson and Giulia Ranzini.
\newblock Click here to look clever: Self-presentation via selective sharing of
  music and film on social media.
\newblock \emph{Computers in Human Behavior}, 82:\penalty0 148--158, 2018.
\newblock \doi{10.1016/j.chb.2018.01.008}.

\bibitem[Johnson and Van Der~Heide(2015)]{Johnson2015}
Benjamin~K. Johnson and Brandon Van Der~Heide.
\newblock Can sharing affect liking? {O}nline taste performances, feedback, and
  subsequent media preferences.
\newblock \emph{Computers in Human Behavior}, 46:\penalty0 181--190, 2015.
\newblock \doi{10.1016/j.chb.2015.01.018}.

\bibitem[Koski(2011)]{Koski2011}
Genevieve Koski.
\newblock Righteous indignation, thy name is {Who Is Arcade Fire??!!?}
\newblock The A.V. Club,
  \url{https://www.avclub.com/righteous-indignation-thy-name-is-who-is-arcade-fire-1798224650},
  February 2011.

\bibitem[Kov{\'a}cs and Sharkey(2014)]{Kovacs2014}
Bal{\'a}zs Kov{\'a}cs and Amanda~J. Sharkey.
\newblock The paradox of publicity: How awards can negatively affect the
  evaluation of quality.
\newblock \emph{Administrative Science Quarterly}, 59\penalty0 (1):\penalty0
  1--33, 2014.
\newblock \doi{10.1177/0001839214523602}.

\bibitem[Leibenstein(1950)]{Leibenstein1950}
Harvey Leibenstein.
\newblock Bandwagon, snob, and {V}eblen effects in the theory of consumers'
  demand.
\newblock \emph{The Quarterly Journal of Economics}, 64\penalty0 (2):\penalty0
  183--207, 1950.
\newblock \doi{10.2307/1882692}.

\bibitem[Lew and Flanagin(2024)]{Lew2024}
Zijian Lew and Andrew~J. Flanagin.
\newblock Self-effects and public commitment on social media: Testing the
  cognitive and social influences of sending messages on message senders.
\newblock \emph{Computers in Human Behavior}, 156:\penalty0 108200, 2024.
\newblock \doi{10.1016/j.chb.2024.108200}.

\bibitem[Liu(2007)]{Liu2007}
Hugo Liu.
\newblock Social network profiles as taste performances.
\newblock \emph{Journal of Computer-Mediated Communication}, 13\penalty0
  (1):\penalty0 252--275, 2007.
\newblock \doi{10.1111/j.1083-6101.2007.00395.x}.

\bibitem[{MetaBrainz Foundation}(2023)]{MLHDplus2023}
{MetaBrainz Foundation}.
\newblock {MLHD+}: The {Music Listening Histories Dataset}, cleaned and linked
  to {MusicBrainz} (release of {March} 2023).
\newblock
  \url{https://data.musicbrainz.org/pub/musicbrainz/listenbrainz/mlhd/};
  documentation at \url{https://musicbrainz.org/doc/MLHD+}, 2023.

\bibitem[{MetaBrainz Foundation}(2026)]{MusicBrainz2026}
{MetaBrainz Foundation}.
\newblock {MusicBrainz} database, full export of 14 {November} 2026.
\newblock \url{https://data.metabrainz.org/pub/musicbrainz/data/fullexport/},
  2026.

\bibitem[Rentfrow and Gosling(2003)]{Rentfrow2003}
Peter~J. Rentfrow and Samuel~D. Gosling.
\newblock The do re mi's of everyday life: The structure and personality
  correlates of music preferences.
\newblock \emph{Journal of Personality and Social Psychology}, 84\penalty0
  (6):\penalty0 1236--1256, 2003.
\newblock \doi{10.1037/0022-3514.84.6.1236}.

\bibitem[Rentfrow and Gosling(2006)]{Rentfrow2006}
Peter~J. Rentfrow and Samuel~D. Gosling.
\newblock Message in a ballad: The role of music preferences in interpersonal
  perception.
\newblock \emph{Psychological Science}, 17\penalty0 (3):\penalty0 236--242,
  2006.
\newblock \doi{10.1111/j.1467-9280.2006.01691.x}.

\bibitem[Shi et~al.(2023)Shi, Liu, Lee, and Srinivasan]{Shi2023}
Zijun~(June) Shi, Xiao Liu, Dokyun Lee, and Kannan Srinivasan.
\newblock How do fast-fashion copycats affect the popularity of premium brands?
  {E}vidence from social media.
\newblock \emph{Journal of Marketing Research}, 60\penalty0 (6):\penalty0
  1027--1051, 2023.
\newblock \doi{10.1177/00222437231164403}.

\bibitem[Silfverberg et~al.(2011)Silfverberg, Liikkanen, and
  Lampinen]{Silfverberg2011}
Suvi Silfverberg, Lassi~A. Liikkanen, and Airi Lampinen.
\newblock ``{I}'ll press play, but {I} won't listen'': Profile work in a
  music-focused social network service.
\newblock In \emph{Proceedings of the ACM 2011 Conference on Computer Supported
  Cooperative Work}, pages 207--216. ACM, 2011.
\newblock \doi{10.1145/1958824.1958855}.

\bibitem[Sim et~al.(2022)Sim, Park, Cho, Smith, and Jung]{Sim2022}
Jaeung Sim, Jea~Gon Park, Daegon Cho, Michael~D. Smith, and Jaemin Jung.
\newblock Bestseller lists and product discovery in the subscription-based
  market: Evidence from music streaming.
\newblock \emph{Journal of Economic Behavior \& Organization}, 194:\penalty0
  550--567, 2022.
\newblock \doi{10.1016/j.jebo.2021.12.030}.

\bibitem[Snyder and Fromkin(1977)]{Snyder1977}
C.~R. Snyder and Howard~L. Fromkin.
\newblock Abnormality as a positive characteristic: The development and
  validation of a scale measuring need for uniqueness.
\newblock \emph{Journal of Abnormal Psychology}, 86\penalty0 (5):\penalty0
  518--527, 1977.
\newblock \doi{10.1037/0021-843X.86.5.518}.

\bibitem[Ternovski and Yasseri(2020)]{Ternovski2020}
John Ternovski and Taha Yasseri.
\newblock Social complex contagion in music listenership: A natural experiment
  with 1.3 million participants.
\newblock \emph{Social Networks}, 61:\penalty0 144--152, 2020.
\newblock \doi{10.1016/j.socnet.2019.10.005}.

\bibitem[Tian et~al.(2001)Tian, Bearden, and Hunter]{Tian2001}
Kelly~Tepper Tian, William~O. Bearden, and Gary~L. Hunter.
\newblock Consumers' need for uniqueness: Scale development and validation.
\newblock \emph{Journal of Consumer Research}, 28\penalty0 (1):\penalty0
  50--66, 2001.
\newblock \doi{10.1086/321947}.

\bibitem[Uski and Lampinen(2016)]{Uski2016}
Suvi Uski and Airi Lampinen.
\newblock Social norms and self-presentation on social network sites: Profile
  work in action.
\newblock \emph{New Media \& Society}, 18\penalty0 (3):\penalty0 447--464,
  2016.
\newblock \doi{10.1177/1461444814543164}.

\bibitem[Vigliensoni and Fujinaga(2017)]{Vigliensoni2017}
Gabriel Vigliensoni and Ichiro Fujinaga.
\newblock The {Music Listening Histories Dataset}.
\newblock In \emph{Proceedings of the 18th International Society for Music
  Information Retrieval Conference}, pages 96--102, Suzhou, China, 2017.
\newblock \doi{10.5281/zenodo.1417499}.

\bibitem[Waddell and Sundar(2020)]{Waddell2020}
T.~Franklin Waddell and S.~Shyam Sundar.
\newblock Bandwagon effects in social television: How audience metrics related
  to size and opinion affect the enjoyment of digital media.
\newblock \emph{Computers in Human Behavior}, 107:\penalty0 106270, 2020.
\newblock \doi{10.1016/j.chb.2020.106270}.

\bibitem[Walsh(2024)]{Walsh2024}
Michael~James Walsh.
\newblock `{I}'m also slightly conscious of how much {I}'m listening to
  something': Music streaming and the transformation of music listening.
\newblock \emph{Media, Culture \& Society}, 46\penalty0 (4):\penalty0 745--761,
  2024.
\newblock \doi{10.1177/01634437231219142}.

\bibitem[Webster(2023)]{Webster2023}
Jack Webster.
\newblock The promise of personalisation: Exploring how music streaming
  platforms are shaping the performance of class identities and distinction.
\newblock \emph{New Media \& Society}, 25\penalty0 (8):\penalty0 2140--2162,
  2023.
\newblock \doi{10.1177/14614448211027863}.

\end{thebibliography}

\appendix
\section{Data construction}
\label{app:data}

\paragraph{Users and plays.} MLHD+ distributes one compressed file per user, split into sixteen archives by user identifier. We read all \nShards ``complete'' archives. Plays dated before January 2005 or after December 2013 are dropped. A play credited to several artists is assigned to the first credited artist. The \nPlaysAllB billion plays in the window come from \nUsersActive users.

\paragraph{The artist universe.} User--artist--month counts are kept for \nUniverse artists: the 10,000 artists with the most listeners in one of the sixteen archives (a random sixteenth of users), every artist that was among the 1,000 most-listened in any single year, and every award nominee. These artists account for most plays; artists outside the universe still count towards a user's total plays, distinct artists and first-time artists.

\paragraph{Newcomers.} A newcomer to an artist in month $m$ is a user whose first play of the artist in the data falls in $m$ and who had played anything at least six months before $m$. The six-month rule keeps users who had just joined Last.fm, whose whole library appears to be new, out of the newcomer counts. A newcomer is \emph{retained} if they played the artist in any month from $m+3$ to $m+12$; retention is computed among newcomers who were still active on Last.fm in that period. A newcomer's mainstream share is the share of their plays in the six months before $m$ that went to that year's 1,000 most-listened artists.

\paragraph{Releases and catalogue.} Release groups in MusicBrainz collect all editions of an album, single or EP. A studio album is a release group of primary type \emph{Album} with no secondary type (compilation, live, soundtrack, remix and so on). The first release date of a group is the earliest date of any of its releases; when the earliest year has no release with a known month, the date has year precision. MLHD+ links each play to a canonical release of its recording, which is usually, though not always, the album on which the recording first appeared; the catalogue measures classify plays by that release rather than by the edition a listener happened to play. Plays are split into the artist's latest studio album, its earlier studio albums, and everything else. In the histories of \CatUsers users from one archive, \CatAlbumPct\% of the plays of the Study~1 treated artists in months $-12$ to $12$ were linked to a studio album credited to the artist. Another \CatOwnOtherPct\% were linked to another release credited to the artist (a single, EP, live album, compilation or soundtrack) and \CatOtherArtistPct\% to a release credited to another artist, such as a various-artists soundtrack or compilation. Every linked release was found in the MusicBrainz database. The last two groups form the residual category. The catalogue results use only artist-events whose latest studio album was the same from month $-12$ to month $12$, so the categories do not change during the comparison.

\paragraph{Release audit.} For every Study~1 artist-event, treated and comparison, we found every MusicBrainz recording credited to the artist in any position, on the recording or on a track, including various-artists releases such as soundtracks, and dated each recording by its earliest release. An artist-event is flagged if any recording first appeared in months $T-3$ to $T+1$, or might have because its earliest date is recorded to the year only. Of the \CrEvents treated events, \AuditNewT (\AuditNewPctT\%) have at least one recording first released in months $T-3$ to $T+1$, and so do \AuditNewC of the \AuditUnitsC comparison artist-events (\AuditNewPctC\%); another \AuditUncT treated and \AuditUncC comparison artist-events have one whose year-only date may fall in that window. In all, \AuditFlaggedT treated and \AuditFlaggedC comparison artist-events are flagged (\AuditFlagPctT\% and \AuditFlagPctC\%). Most involve one or two recordings: \AuditBatchT treated and \AuditBatchC comparison artist-events had three or more recordings with a month-precise date in the window. A re-release of an existing recording, such as an older song on a new soundtrack, is not flagged. The flags rest on MusicBrainz metadata and mark possible rather than verified new material. Of the \AuditCleanT treated events without a flag, \AuditCleanNoCtrl lose all their comparison artists once flagged comparison artists are dropped, which leaves \AuditEvents events in the audited analysis.

\paragraph{Album histories.} MusicBrainz classifies release groups by type, and some albums in it were released only as promotional copies. Kavinsky's \emph{Dead Cruiser} (2007), for example, has no official release, so the main specification counts it as a studio album, although the artist's first official album appeared in 2013. A sensitivity analysis counts a studio album only if at least one of its releases is official; this changes the album histories of \OfficialChanged treated artists, and Kavinsky, who then has no earlier album, drops out (Appendix~\ref{app:robust}).

\paragraph{Award nominees.} Nominees and winners come from the Wikipedia pages of the three Grammy categories and of the Mercury Prize. A losing nominee is used as a comparison only if they won no Grammy at all that night, as listed on the Wikipedia page of the ceremony; packaging, remix and producer awards, which go to art directors, remixers and producers, do not count as wins for the performing artist. Ceremony dates come from the Wikipedia pages of each Grammy ceremony and from press reports of each Mercury ceremony. Names were matched to MusicBrainz artists, preferring the match with the most listeners; three artists renamed after 2013 (British Sea Power, Kanye West, Lady Antebellum) were matched under their current names, and duo albums were assigned to the first credited artist, as in the listening data.

\paragraph{Sample flow.} Table~\ref{tab:flow} traces both samples from the full data to the analysed events. In Study~2, the Mercury Prize of 2009 drops out because its winner had fewer than 30 early fans.

\begin{table}[H]
\centering
\small
\caption{Sample flow.}
\label{tab:flow}
\begin{tabular}{p{0.72\textwidth}r}
\toprule
Step & Count \\
\midrule
\emph{Study 1: crowd arrivals} &  \\
Artist-months meeting the surge and audience rules, 2007--2012 & 3,130 \\
\quad with no new release around the event & 246 \\
\quad and no death or break-up & 243 \\
\quad and no surge a year before or after (not seasonal) & 165 \\
Artists, keeping each artist's largest surge & 101 \\
\quad with at least one matched comparison artist & 78 \\
\quad with $\geq$30 early fans and a comparison with $\geq$30 (analysed) & 78 \\
Comparison artist-events analysed & 357 \\
\emph{Study 2: award wins} &  \\
Nominee slots (Grammy 2008--2012, three categories; Mercury 2007--2011) & 135 \\
Ceremony--artist units & 131 \\
\quad excluding nominees who won another Grammy that night & 116 \\
Winner artist-events analysed ($\geq$30 early fans) & 18 \\
Losing-nominee artist-events analysed & 80 \\
Ceremonies analysed & 9 \\
\bottomrule
\end{tabular}
 
\end{table}

\section{Comparability of treated and comparison artists}
\label{app:balance}

\begin{table}[H]
\centering
\small
\caption{Study 1: treated and comparison artist-events before the event. Means or medians across artist-events. As matched: each event's comparison artists share one unit of weight, as in the estimator; unweighted: all comparison artist-events count equally.}
\label{tab:balance}
\begin{tabular}{p{0.52\textwidth}rrr}
\toprule
 & Treated & \multicolumn{2}{c}{Comparison} \\
\cmidrule(lr){3-4}
 & & As matched & Unweighted \\
\midrule
Early fans per artist (median) & 846 & 1,005 & 952 \\
Monthly listeners, months $-12$ to $-7$ (median) & 614 & 810 & 760 \\
Newcomers, months $-6$ to $-1$ vs $-12$ to $-7$ (median ratio) & 1.28 & 1.13 & 1.11 \\
Artist's share of early fans' plays, months $-12$ to $-7$ (\%) & 0.44 & 0.37 & 0.36 \\
Months early fans played the artist, months $-12$ to $-7$ (\%) & 39 & 40 & 39 \\
Pre-event slope of the share, months $-6$ to $-1$ (pp per month) & $-$0.0174 & $-$0.0130 & $-$0.0126 \\
Early fans' tenure with the artist (median months) & 20 & 21 & 21 \\
Early fans' mainstream share of other listening (\%) & 50 & 48 & 47 \\
Distinct months with plays in the selection window (mean) & 5.7 & 6.0 & 6.0 \\
Months since latest studio album (median) & 34 & 34 & 31 \\
Early fans who stopped playing the artist, months $-12..{-7}$ to $-6..{-1}$ (\%) & 28.5 & 28.6 & 29.0 \\
Lapsed early fans who came back, months $-12..{-7}$ to $-6..{-1}$ (\%) & 26.8 & 27.3 & 26.8 \\
\midrule
Artist-events & 78 & \multicolumn{2}{c}{357} \\
\bottomrule
\end{tabular}
 
\end{table}

Table~\ref{tab:balance} compares the two groups before the event. Treated and comparison artists are similar in their early fans' listening, tenure and taste, in the pre-event slope of the share of plays, and in how often early fans stopped playing the artist or came back in the months before the event. Weighting comparison artists as in the estimator changes little. Treated artists had smaller audiences and, as Figure~\ref{fig:crowd}A shows, newcomers had already begun to rise in the months before the surge. Of the \CrControlUnits comparison artist-events, \CrControlArtists are distinct artists, and \CrBothRoles artist appears as treated in one event and as a comparison in another.

\section{Outcome samples}
\label{app:samples}

Table~\ref{tab:outcomes} gives the samples behind each Study~1 outcome. A fan unit is one early fan of one artist-event; a user can be an early fan of several artists. The quitting and returning comparisons use artist-events with at least ten eligible fans and fans who used Last.fm in at least four months of each six-month window, so they describe fans who stayed observable on the platform. Fewer than four active months in months $0$ to $5$ were recorded for \CrLowObsPctT\% of treated and \CrLowObsPctC\% of comparison fan units.

\begin{table}[H]
\centering
\small
\caption{Study 1 outcome samples: treated and comparison artist-events, fan units and distinct users, and the units at risk in each transition.}
\label{tab:outcomes}
\begin{tabular}{p{0.62\textwidth}rr}
\toprule
 & Treated & Comparison \\
\midrule
Artist-events analysed & 78 & 357 \\
Early-fan units (fan $\times$ artist-event) & 102,617 & 589,810 \\
Distinct users & 75,200 & 209,676 \\
\emph{Primary endpoint} (artist-events with $\geq$10 eligible fans) &  &  \\
Quitting: played in months $-6..{-1}$, $\geq$4 active months in $-6..{-1}$ and $0..5$ & 63,633 & 373,871 \\
Returning: no plays in months $-6..{-1}$, same activity rule & 34,745 & 188,501 \\
Quitting placebo: played in $-12..{-7}$, $\geq$4 active months in $-12..{-7}$ and $-6..{-1}$ & 80,666 & 466,912 \\
Returning placebo: no plays in $-12..{-7}$, same activity rule & 19,481 & 107,395 \\
Fewer than four active months in $0..5$ (\% of units) & 3.5 & 4.0 \\
Fewer than four active months in $-6..{-1}$ (\% of units) & 0.9 & 1.0 \\
\emph{Secondary outcomes} &  &  \\
Listened at all and share of plays: units with any activity in the window & 102,617 & 589,810 \\
Relative share: units with plays of the artist in months $-12..{-7}$ & 82,688 & 479,644 \\
\bottomrule
\end{tabular}
 
\end{table}

\section{Study 1 event ledger}
\label{app:events}

Table~\ref{tab:ledger} lists every crowd arrival analysed in Study~1. The detector knows nothing about why an audience grew. For \LedgerDocumented events we verified the exposure that coincides with the surge against the Wikipedia pages of the film, soundtrack or episode; the others are listed without a documented source. Fans: early fans before sampling.

{\footnotesize
\setlength{\tabcolsep}{4pt}
\begin{longtable}{p{3.2cm}lrrrrlp{3.4cm}}
\caption{Study 1 event ledger: every crowd arrival analysed. Surge: newcomers in months 0--2 relative to the previous year's monthly mean (three-month average over the monthly mean). Growth: monthly listeners in months 0--5 relative to months $-12$ to $-7$. Album: months since the latest studio album in MusicBrainz. New rec.: recordings credited to the artist that first appeared in months $-3$ to $+1$ (yes), possibly did because the date is year-only (?), or did not (no). Documented exposures were checked against the English Wikipedia articles on the film \emph{Drive} and its soundtrack, the \emph{Twilight} and \emph{New Moon} soundtrack albums, the \emph{Family Guy} episode ``I Dream of Jesus'' and the film \emph{The Muppets}.}\label{tab:ledger}\\
\toprule
Artist & Month & Surge & Growth & Album & Fans & New rec. & Documented exposure \\
\midrule
\endfirsthead
\toprule
Artist & Month & Surge & Growth & Album & Fans & New rec. & Documented exposure \\
\midrule
\endhead
\bottomrule
\endfoot
Nokturnal Mortum & 2007-08 & 3.1$\times$ & 113\% & 37 & 519 & no &  \\
Secret Machines & 2007-09 & 3.1$\times$ & 54\% & 17 & 2,494 & yes &  \\
Stephen Malkmus & 2007-11 & 3.3$\times$ & 105\% & 30 & 2,245 & yes &  \\
Bell X1 & 2008-01 & 3.0$\times$ & 87\% & 36 & 1,143 & yes &  \\
JamisonParker & 2008-01 & 3.4$\times$ & 48\% & 30 & 1,201 & no &  \\
Sonic Youth & 2008-01 & 3.2$\times$ & 88\% & 19 & 15,438 & yes &  \\
Camel & 2008-02 & 3.3$\times$ & 136\% & 67 & 1,665 & yes &  \\
Pati Yang & 2008-02 & 3.6$\times$ & 314\% & 28 & 562 & ? &  \\
The Wohlstandskinder & 2008-02 & 3.2$\times$ & 81\% & 49 & 553 & no &  \\
Admiral Freebee & 2008-03 & 3.3$\times$ & 82\% & 17 & 574 & no &  \\
Hanne Hukkelberg & 2008-03 & 3.2$\times$ & 114\% & 18 & 1,320 & ? &  \\
Jackson and His Computer Band & 2008-03 & 3.3$\times$ & 74\% & 44 & 819 & yes &  \\
Slash's Snakepit & 2008-03 & 4.1$\times$ & 163\% & 90 & 512 & no &  \\
Timo Rautiainen \& Trio Niskalaukaus & 2008-03 & 3.1$\times$ & 70\% & 42 & 934 & no &  \\
Cazuza & 2008-05 & 3.2$\times$ & 163\% & 118 & 513 & no &  \\
Greenskeepers & 2008-05 & 4.5$\times$ & 150\% & 7 & 338 & yes &  \\
Fingathing & 2008-06 & 5.1$\times$ & 189\% & 49 & 599 & no &  \\
Elefant & 2008-07 & 3.8$\times$ & 97\% & 27 & 2,412 & no &  \\
Freedom Call & 2008-07 & 5.3$\times$ & 103\% & 15 & 1,521 & no &  \\
Kubb & 2008-07 & 4.0$\times$ & 118\% & 32 & 581 & ? &  \\
Moondog & 2008-07 & 3.0$\times$ & 184\% & 156 & 745 & yes &  \\
Sir Mix-A-Lot & 2008-07 & 3.5$\times$ & 130\% & 58 & 247 & yes &  \\
Surkin & 2008-07 & 3.0$\times$ & 654\% & 17 & 348 & yes &  \\
The Like & 2008-07 & 3.6$\times$ & 91\% & 34 & 917 & no &  \\
The Sound & 2008-07 & 3.0$\times$ & 212\% & 111 & 589 & no &  \\
The Velvet Teen & 2008-07 & 3.2$\times$ & 67\% & 25 & 941 & no &  \\
Mahogany & 2008-08 & 3.4$\times$ & 118\% & 22 & 636 & no &  \\
The Bluetones & 2008-08 & 3.5$\times$ & 157\% & 22 & 1,097 & no &  \\
Virgin Steele & 2008-08 & 5.3$\times$ & 128\% & 23 & 661 & yes &  \\
Shugo Tokumaru & 2008-08 & 3.9$\times$ & 231\% & 10 & 747 & yes &  \\
Bratmobile & 2008-09 & 3.9$\times$ & 111\% & 74 & 546 & no &  \\
Satanic Warmaster & 2008-10 & 3.6$\times$ & 109\% & 27 & 653 & yes &  \\
The Mercury Program & 2008-10 & 5.4$\times$ & 125\% & 73 & 765 & no &  \\
The Trashmen & 2008-10 & 3.8$\times$ & 178\% & 531 & 517 & no & \emph{Family Guy} episode (Oct.\ 2008) \\
Blue Foundation & 2008-11 & 8.9$\times$ & 564\% & 14 & 1,083 & ? & \emph{Twilight} soundtrack (Nov.\ 2008) \\
Farin Urlaub & 2008-12 & 3.6$\times$ & 66\% & 45 & 3,292 & no &  \\
22-20s & 2009-01 & 3.2$\times$ & 111\% & 57 & 968 & no &  \\
Black Strobe & 2009-01 & 3.7$\times$ & 148\% & 20 & 857 & yes &  \\
Café Tacvba & 2009-01 & 4.6$\times$ & 163\% & 17 & 423 & yes &  \\
Millionaire & 2009-01 & 3.6$\times$ & 48\% & 43 & 1,009 & no &  \\
Kevin Drew & 2009-02 & 6.6$\times$ & 160\% & 18 & 1,776 & yes &  \\
Amduscia & 2009-03 & 3.1$\times$ & 106\% & 14 & 758 & no &  \\
Lavender Diamond & 2009-03 & 3.4$\times$ & 50\% & 22 & 1,364 & yes &  \\
Jr. Walker \& The All Stars & 2009-04 & 3.0$\times$ & 232\% & 177 & 248 & yes &  \\
Skeeter Davis & 2009-04 & 4.5$\times$ & 253\% & 213 & 273 & yes &  \\
The Tremeloes & 2009-04 & 3.3$\times$ & 190\% & 105 & 312 & ? &  \\
Mumm-Ra & 2009-10 & 3.5$\times$ & 228\% & 29 & 2,104 & no &  \\
Anya Marina & 2009-11 & 4.9$\times$ & 610\% & 12 & 797 & yes & \emph{New Moon} soundtrack (Oct.\ 2009) \\
The Snake the Cross the Crown & 2009-12 & 3.1$\times$ & 35\% & 33 & 1,280 & no &  \\
Ryan Bingham & 2010-03 & 3.5$\times$ & 318\% & 29 & 450 & yes &  \\
The Dickies & 2010-04 & 6.7$\times$ & 170\% & 107 & 1,001 & yes &  \\
Zongamin & 2010-04 & 4.2$\times$ & 73\% & 85 & 834 & yes &  \\
Elf & 2010-05 & 3.1$\times$ & 138\% & 46 & 474 & no &  \\
FM Belfast & 2010-05 & 3.1$\times$ & 342\% & 19 & 772 & yes &  \\
Radical Face & 2010-07 & 3.0$\times$ & 177\% & 40 & 2,463 & no &  \\
Dan the Automator & 2010-08 & 4.1$\times$ & 117\% & 121 & 330 & yes &  \\
Frank Black & 2010-08 & 5.3$\times$ & 122\% & 50 & 4,584 & ? &  \\
The Fireman & 2010-09 & 3.1$\times$ & 47\% & 22 & 2,383 & no &  \\
Two Fingers & 2010-09 & 7.5$\times$ & 236\% & 12 & 806 & yes &  \\
Artist Vs Poet & 2010-10 & 3.1$\times$ & 131\% & 7 & 1,901 & yes &  \\
The Little Willies & 2010-11 & 11.2$\times$ & 219\% & 56 & 976 & no &  \\
The Phoenix Foundation & 2010-11 & 3.4$\times$ & 229\% & 7 & 661 & no &  \\
Christian Walz & 2011-02 & 3.1$\times$ & 57\% & 27 & 708 & yes &  \\
Marcelo D2 & 2011-05 & 3.3$\times$ & 64\% & 9 & 1,731 & yes &  \\
Brooke Fraser & 2011-06 & 4.2$\times$ & 241\% & 8 & 1,581 & yes &  \\
Graham Nash & 2011-06 & 4.4$\times$ & 88\% & 110 & 922 & yes &  \\
Desire & 2011-09 & 45.4$\times$ & 2,086\% & 30 & 858 & yes & Film \emph{Drive} (Sept.\ 2011) \\
Kavinsky & 2011-09 & 12.9$\times$ & 504\% & 56 & 3,520 & ? & Film \emph{Drive} (Sept.\ 2011), opening song \\
Felt & 2011-10 & 4.2$\times$ & 68\% & 263 & 2,260 & yes &  \\
Gavin Friday & 2011-10 & 3.0$\times$ & 86\% & 7 & 541 & yes &  \\
The Chills & 2011-10 & 4.2$\times$ & 73\% & 232 & 1,393 & yes &  \\
Keith Murray & 2011-11 & 4.0$\times$ & 132\% & 42 & 668 & yes &  \\
The Muppets & 2011-11 & 11.9$\times$ & 325\% & 70 & 339 & yes & Film \emph{The Muppets} (Nov.\ 2011) \\
Liturgy & 2011-12 & 3.7$\times$ & 257\% & 7 & 1,351 & no &  \\
Farin Urlaub Racing Team & 2012-02 & 4.2$\times$ & 2\% & 40 & 2,560 & no &  \\
Faunts & 2012-03 & 3.8$\times$ & 56\% & 37 & 2,267 & yes &  \\
Joe Pass & 2012-03 & 3.1$\times$ & 37\% & 87 & 1,711 & yes &  \\
Marie Laforêt & 2012-03 & 4.7$\times$ & 76\% & 173 & 676 & yes &  \\
\end{longtable}
 
}

\section{Study 2: award winners}
\label{app:awards}

\begin{figure}[H]
\centering
\includegraphics[width=\textwidth]{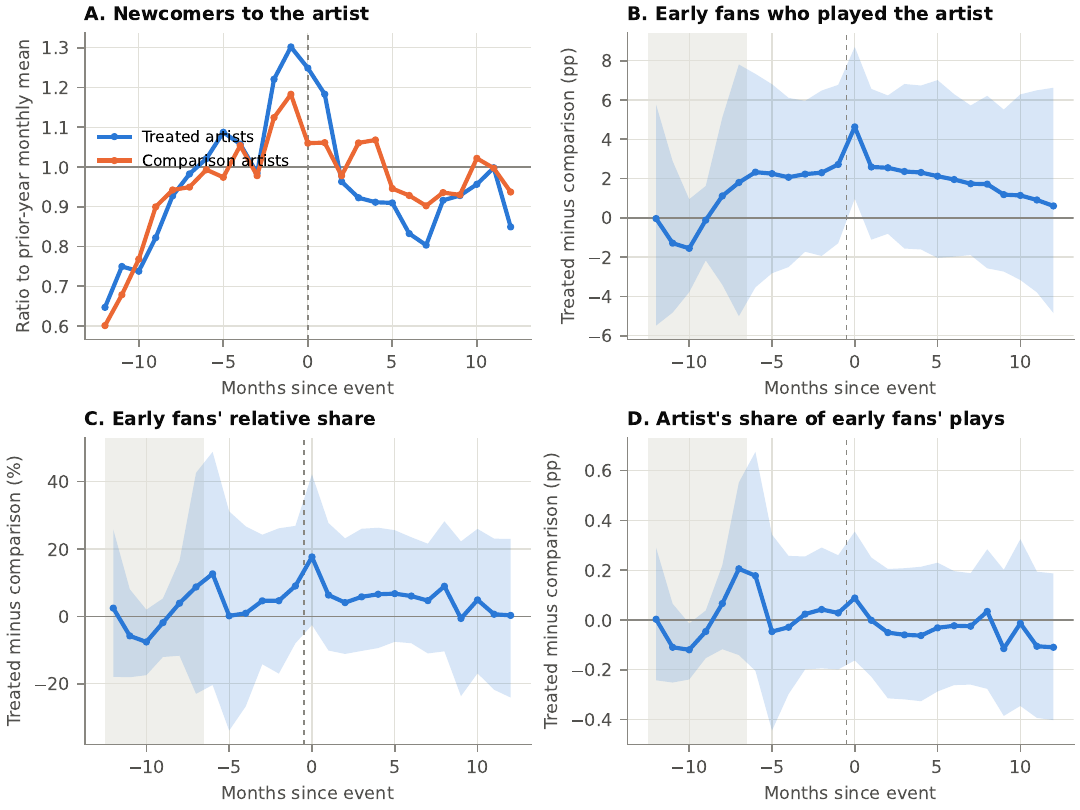}
\caption{Study 2, award wins, by month. Panel A: newcomers to winners and to losing nominees, as the median ratio to each artist's monthly mean in the year before the ceremony. Panels B--D: change in early fans' outcomes for winners minus the same ceremony's losing nominees, relative to months $-12$ to $-7$, with 95\% bootstrap intervals over ceremonies.}
\label{fig:awards}
\end{figure}

Table~\ref{tab:winners} lists the \AwNevents winner artist-events of Study~2. Few winners drew many more newcomers than in the previous year; only Esperanza Spalding, Best New Artist in 2011, drew more than twice as many. The change in early fans' relative share, measured against the same ceremony's losing nominees, varies widely from winner to winner and in both directions, which is why the pooled estimate is imprecise. Figure~\ref{fig:awards} shows the monthly paths. Newcomers to winners and to losing nominees rose together over the year before the ceremony, as the nominated albums came out, and around the ceremony winners' newcomers peaked only slightly higher. The differences in early fans' listening move within wide intervals that mostly include zero.

\begin{table}[H]
\centering
\small
\caption{Award winners in Study 2. Won: categories the artist won among the three Grammy categories or the Mercury Prize. Newcomer ratio: newcomers in the ceremony month and the two following months relative to the previous year's monthly mean. Relative share: change in early fans' relative share minus the mean change for the same ceremony's losing nominees.}
\label{tab:winners}
\begin{tabular}{lllrrr}
\toprule
Ceremony & Winner & Won & Newcomer ratio & Early fans & Relative share (\%) \\
\midrule
Grammy 2008 & Amy Winehouse & BNA & 1.36 & 11,171 & $-$28.4 \\
Grammy 2008 & Herbie Hancock & AOTY & 1.54 & 3,384 & $-$6.6 \\
Grammy 2008 & The White Stripes & ALT & 0.83 & 34,823 & $-$21.3 \\
Grammy 2009 & Adele & BNA & 1.15 & 7,025 & $-$42.9 \\
Grammy 2009 & Radiohead & ALT & 1.29 & 93,390 & 18.5 \\
Grammy 2009 & Robert Plant & AOTY & 1.76 & 6,003 & $-$28.0 \\
Grammy 2010 & Phoenix & ALT & 1.01 & 29,155 & 10.4 \\
Grammy 2010 & Taylor Swift & AOTY & 0.77 & 13,942 & 22.4 \\
Grammy 2010 & Zac Brown Band & BNA & 1.82 & 514 & 20.8 \\
Grammy 2011 & Arcade Fire & AOTY & 1.26 & 79,945 & 61.4 \\
Grammy 2011 & Esperanza Spalding & BNA & 3.82 & 1,734 & $-$1.9 \\
Grammy 2011 & The Black Keys & ALT & 1.00 & 37,686 & 35.6 \\
Grammy 2012 & Adele & AOTY & 0.58 & 90,295 & $-$11.7 \\
Grammy 2012 & Bon Iver & ALT, BNA & 1.00 & 83,802 & 19.1 \\
Mercury 2007 & Klaxons & MERCURY & 1.24 & 4,015 & $-$10.4 \\
Mercury 2008 & Elbow & MERCURY & 1.75 & 5,203 & 86.1 \\
Mercury 2010 & The xx & MERCURY & 0.75 & 49,866 & $-$31.3 \\
Mercury 2011 & PJ Harvey & MERCURY & 0.87 & 47,087 & 6.9 \\
\bottomrule
\end{tabular}
 
\end{table}

\section{Robustness}
\label{app:robust}

Table~\ref{tab:robust} repeats the main Study~1 estimates under alternative choices. They include a fan-weighted rather than artist-weighted average, stricter definitions of an early fan (the extraction keeps user--artist pairs with at least ten lifetime plays, so looser definitions are not possible), and lower and higher surge thresholds. Another variant chooses comparison artists only on information up to month $-1$: no surge starting after month $-3$, since a surge ratio uses three months of newcomers, and no album, death or break-up before the event. It is reported for the same events and for all events that this rule allows to be matched. Further variants drop every artist-event, treated or comparison, with any new recording near the event, or with three or more, alone and combined with the pre-event comparison rule. The remaining variants count a studio album only if it was officially released, drop events in November to January, when holiday listening and year-end lists change listening patterns, and drop events before 2009, when the user base was smaller. Quitting falls in all \RobVariants alternative specifications (from \RobQuitMin to \RobQuitMax points), and none of the 95\% intervals includes zero. The share of months with any listening rises in all of them. The relative share and the share of plays depend more on the choice of comparison artists. Chosen on pre-event information only, \PastCtrlAlbPct\% of the comparison artists released a studio album and \PastCtrlSurgePct\% had a surge month of their own within the year after the event, which can raise their fans' listening. With these comparisons the relative share no longer rises clearly and the share of plays falls more.

\begin{table}[H]
\centering
\small
\caption{Study 1 under alternative specifications. Quit: treated minus comparison difference in the share of early fans who stopped playing the artist (months $-6..{-1}$ to $0..5$). Other columns: treated minus comparison change from months $-12$ to $-7$ to months $0$ to $12$. 95\% bootstrap intervals over events.}
\label{tab:robust}
\resizebox{\textwidth}{!}{\begin{tabular}{lcccc r}
\toprule
Specification & Quit (pp) & Share of plays (pp) & Relative share (\%) & Listened (pp) & Events \\
\midrule
Main specification & $-$3.3 [$-$4.6, $-$1.9] & $-$0.024 [$-$0.047, $-$0.002] & 6.2 [3.3, 9.1] & 2.8 [2.1, 3.4] & 78 \\
Fans weighted equally (not artist-events) & $-$3.3 [$-$4.4, $-$2.1] & 0.005 [$-$0.034, 0.037] & 7.3 [3.5, 10.5] & 2.8 [1.9, 3.6] & 78 \\
Early fans: $\geq$15 plays & $-$3.0 [$-$4.2, $-$1.8] & $-$0.026 [$-$0.053, 0.001] & 5.7 [2.8, 8.7] & 3.1 [2.4, 3.8] & 78 \\
Early fans: $\geq$25 plays & $-$2.5 [$-$3.8, $-$1.1] & $-$0.026 [$-$0.059, 0.005] & 5.3 [2.3, 8.4] & 3.3 [2.5, 4.2] & 78 \\
Surge threshold 2$\times$ & $-$2.3 [$-$2.8, $-$1.9] & $-$0.008 [$-$0.020, 0.004] & 5.3 [4.1, 6.4] & 2.7 [2.4, 3.0] & 512 \\
Surge threshold 5$\times$ & $-$5.7 [$-$7.8, $-$3.3] & $-$0.032 [$-$0.085, 0.019] & 11.3 [5.5, 17.8] & 3.9 [3.0, 5.1] & 13 \\
Comparisons chosen with information before the event only (same events) & $-$2.9 [$-$4.1, $-$1.7] & $-$0.050 [$-$0.073, $-$0.026] & 2.7 [$-$0.2, 5.6] & 1.8 [1.2, 2.5] & 78 \\
Comparisons chosen with information before the event only (all eligible events) & $-$2.6 [$-$3.8, $-$1.4] & $-$0.053 [$-$0.076, $-$0.031] & 1.6 [$-$1.3, 4.6] & 1.5 [0.8, 2.1] & 86 \\
Pre-event comparisons and no new recording near the event (audit) & $-$4.2 [$-$5.8, $-$2.4] & $-$0.065 [$-$0.102, $-$0.031] & 6.1 [1.6, 10.7] & 2.2 [1.1, 3.3] & 27 \\
Studio albums counted only if officially released & $-$3.0 [$-$4.2, $-$1.6] & $-$0.024 [$-$0.046, $-$0.004] & 5.5 [2.5, 8.3] & 2.7 [2.0, 3.4] & 77 \\
No new recording of any kind near the event (audit) & $-$3.4 [$-$5.4, $-$1.2] & $-$0.018 [$-$0.053, 0.016] & 7.4 [2.3, 12.6] & 2.8 [1.8, 3.8] & 27 \\
Fewer than three new recordings near the event (audit) & $-$2.9 [$-$4.3, $-$1.3] & $-$0.024 [$-$0.048, 0.000] & 5.6 [2.2, 8.8] & 2.6 [1.9, 3.3] & 64 \\
Events outside November--January & $-$2.9 [$-$4.3, $-$1.3] & $-$0.027 [$-$0.053, $-$0.003] & 5.4 [2.2, 8.7] & 2.8 [2.0, 3.6] & 61 \\
Events from 2009 onwards & $-$3.3 [$-$5.1, $-$1.4] & $-$0.014 [$-$0.043, 0.012] & 7.0 [2.8, 11.1] & 2.6 [1.6, 3.5] & 42 \\
\bottomrule
\end{tabular}
 }
\end{table}

\paragraph{Dependence between events.} Events in the same calendar month can share an exposure (Kavinsky and Desire both rose with the film \emph{Drive}), and \CrControlArtists distinct comparison artists serve \CrControlUnits comparison artist-events. Resampling the \CrClusterMonths calendar months of the events, rather than the events, gives a 95\% interval for the quitting difference of \CrQuitDiffMonthlo to \CrQuitDiffMonthhi points. Linking events that share a calendar month or any comparison artist yields \CrClusterComps groups, the largest with \CrClusterCompMax events; resampling these groups gives \CrQuitDiffComplo to \CrQuitDiffComphi points.

\paragraph{Fan-level regressions.} The regressions include fan fixed effects, event-by-window fixed effects (windows $-12..{-7}$, $-6..{-1}$ and $0..12$) and weights that give each artist-event equal total weight, with standard errors clustered by artist (\CrRegClusters artists). The post-event coefficient is \CrRegShare\ percentage points (SE \CrRegSharese) on the share of plays, \CrRegAny\ percentage points (SE \CrRegAnyse) on listening at all, and \CrRegRel\% (SE \CrRegRelse) on the relative share. The corresponding pre-event coefficients are \CrRegSharePre\ (SE \CrRegSharePrese), \CrRegAnyPre\ (SE \CrRegAnyPrese) and \CrRegRelPre\ (SE \CrRegRelPrese).

\paragraph{The relative share.} The relative share is defined only for early fans who played the artist in months $-12$ to $-7$; \CrZeroBasePctT\% of treated and \CrZeroBasePctC\% of comparison fan units had no plays then and are excluded from that outcome only. Restricted to fans with plays in that window, the share of plays changed by \CrPosBaseShare points (95\% CI \CrPosBaseSharelo to \CrPosBaseSharehi) and listening at all by \CrPosBaseAny points (95\% CI \CrPosBaseAnylo to \CrPosBaseAnyhi). The cap of ten times a fan's reference share applies to \CrRelCapMonthPct\% of the fan-months after the event, and \CrRelCapUnitPct\% of fan units have at least one capped month. With a cap of five, the relative share rises by \CrRelCapFive\% (95\% CI \CrRelCapFivelo\% to \CrRelCapFivehi\%); with a cap of fifty, by \CrRelCapFifty\% (95\% CI \CrRelCapFiftylo\% to \CrRelCapFiftyhi\%). Early fans of treated and comparison artists used Last.fm in a similar number of months in every window (for example \CrActMonthsPostT and \CrActMonthsPostC of the thirteen months after the event).

\paragraph{The share of plays by tenure.} The small dip in the artist's share of total plays differs by how long early fans had known the artist. For the third who had first played it at most \CrTenCutLo months before the event, the share changed by \CrTenShareA percentage points (95\% CI \CrTenShareAlo to \CrTenShareAhi) from \CrTenShareRefA\%. It changed by \CrTenShareB for the middle third (95\% CI \CrTenShareBlo to \CrTenShareBhi) and by \CrTenShareC for the third who had known it for more than \CrTenCutHi months (95\% CI \CrTenShareClo to \CrTenShareChi). Measured from months $-6$ to $-1$, none of the three changes differs from zero (\CrTenSharePostPreA, \CrTenSharePostPreB and \CrTenSharePostPreC points).

\paragraph{Strict early fans.} Among early fans who had first played the artist at least 24 months before the event and in at least three separate months of the selection window (\CrStrictPctT\% of treated fan units), the quitting difference was \CrStrictQuitDiff points (95\% CI \CrStrictQuitDifflo to \CrStrictQuitDiffhi). Listening at all rose by \CrStrictAny points (95\% CI \CrStrictAnylo to \CrStrictAnyhi) and the relative share by \CrStrictRel\% (95\% CI \CrStrictRello\% to \CrStrictRelhi\%).

\paragraph{Within-fan comparison.} Some early fans of a treated artist were also early fans of one of its comparison artists. For these \CrWithinFans fans, comparing the change in their listening to the two artists holds the person fixed. With each event weighted equally, the treated artist gained \CrWithinAny points more months of listening than the comparison artist (95\% CI \CrWithinAnylo to \CrWithinAnyhi) and \CrWithinRel\% in relative share (95\% CI \CrWithinRello\% to \CrWithinRelhi\%), with no clear difference in the share of plays (\CrWithinShare points, 95\% CI \CrWithinSharelo to \CrWithinSharehi). These estimates point the same way as the main results but are imprecise, because one event accounts for \CrWithinTopPct\% of these fans.

\paragraph{Artists with recent albums.} The results hold for artists whose latest studio album was at most two years old at the time of the surge (\CrAgeRecentN events; relative share \CrAgeRecentRel\%, 95\% CI \CrAgeRecentRello\% to \CrAgeRecentRelhi\%; listened at all \CrAgeRecentAny points, 95\% CI \CrAgeRecentAnylo to \CrAgeRecentAnyhi). They also hold for artists rediscovered for older work (\CrAgeOlderN events; \CrAgeOlderRel\%, 95\% CI \CrAgeOlderRello\% to \CrAgeOlderRelhi\%; \CrAgeOlderAny points, 95\% CI \CrAgeOlderAnylo to \CrAgeOlderAnyhi).

\paragraph{Newcomer cohorts.} Comparing surge arrivals with newcomers from months $-12$ to $-7$, whose follow-up overlaps the event, gives a retention difference of \CrRetNearDiD points (95\% CI \CrRetNearDiDlo to \CrRetNearDiDhi); like the main estimate, it does not differ clearly from zero.

\end{document}